\documentclass[sigconf]{acmart}
\usepackage{tabularx,multirow,multicol}
\usepackage{makecell}

\AtBeginDocument{%
  }

\setcopyright{acmlicensed}
\copyrightyear{2018}
\acmYear{2018}
\acmDOI{XXXXXXX.XXXXXXX}
\acmConference[Conference acronym 'XX]{Make sure to enter the correct
  conference title from your rights confirmation email}{June 03--05,
  2018}{Woodstock, NY}
\acmISBN{978-1-4503-XXXX-X/2018/06}

\begin{document}

\title[User Perceptions on Sensitive Attribute Inference]{Through Their Eyes: User Perceptions on Sensitive Attribute Inference of Social Media Videos by Visual Language Models}


\author{Shuning Zhang}
\orcid{0000-0002-4145-117X}
\email{zsn23@mails.tsinghua.edu.cn}
\affiliation{%
  \institution{Tsinghua University}
  \city{Beijing}
  \country{China}
}

\author{Gengrui Zhang}
\email{zgr23@mails.tsinghua.edu.cn}
\affiliation{%
  \institution{Tsinghua University}
  \city{Beijing}
  \country{China}
}

\author{Yibo Meng}
\email{mengyb22@mails.tsinghua.edu.cn}
\affiliation{%
  \institution{Tsinghua University}
  \city{Beijing}
  \country{China}
}

\author{Ziyi Zhang}
\email{ziyiz13@illinois.edu}
\affiliation{%
  \institution{Southeast University}
  \city{Nanjing}
  \country{China}
}

\author{Hantao Zhao}
\email{htzhao@seu.edu.cn}
\affiliation{%
  \institution{Southeast University}
  \city{Nanjing}
  \country{China}
}

\author{Xin Yi}
\orcid{0000-0001-8041-7962}
\authornote{Corresponding author.}
\email{yixin@tsinghua.edu.cn}
\affiliation{
    \institution{Tsinghua University}
    \city{Beijing}
    \country{China}
}

\author{Hewu Li}
\orcid{0000-0002-6331-6542}
\email{lihewu@cernet.edu.cn}
\affiliation{
    \institution{Tsinghua University}
    \city{Beijing}
    \country{China}
}

\renewcommand{\shortauthors}{Trovato et al.}

\begin{abstract}
  The rapid advancement of Visual Language Models (VLMs) has enabled sophisticated analysis of visual content, leading to concerns about the inference of sensitive user attributes and subsequent privacy risks. While technical capabilities of VLMs are increasingly studied, users' understanding, perceptions, and reactions to these inferences remain less explored, especially concerning videos uploaded on the social media. This paper addresses this gap through a semi-structured interview (N=17), investigating user perspectives on VLM-driven sensitive attribute inference from their visual data. Findings reveal that users perceive VLMs as capable of inferring a range of attributes, including location, demographics, and socioeconomic indicators, often with unsettling accuracy. Key concerns include unauthorized identification, misuse of personal information, pervasive surveillance, and harm from inaccurate inferences. Participants reported employing various mitigation strategies, though with skepticism about their ultimate effectiveness against advanced AI. Users also articulate clear expectations for platforms and regulators, emphasizing the need for enhanced transparency, user control, and proactive privacy safeguards. These insights are crucial for guiding the development of responsible AI systems, effective privacy-enhancing technologies, and informed policymaking that aligns with user expectations and societal values.
\end{abstract}




\received{20 February 2007}
\received[revised]{12 March 2009}
\received[accepted]{5 June 2009}

\maketitle

\section{Introduction}

The proliferation of digital cameras and ubiquitous sharing of visual content on online platforms have created an unprecedented volume of user-generated images and videos~\cite{hua2024generative,zhang2024confrontation}. Concurrently, Visual Language Models (VLMs) have emerged, demonstrating remarkable capabilities in interpreting and generating human-like inferences from visual data by integrating computer vision and natural language processing~\cite{wu2020privacy,orekondy2018connecting,orekondy2017towards}. These models, often pre-trained on vast datasets, can discern subtle visual cues indicative of sensitive attributes ranging from demographic information (age, gender) to more nuanced details like health status~\cite{duddu2022inferring,mehnaz2022your}, religious beliefs, or even personality traits~\cite{nimmo2024user}. Such inferences can be leveraged for constructing detailed personal profiles~\cite{staab2023beyond}, potentially leading to (re)identification or cross-app tracking~\cite{reardon201950,griffiths2018privacy}.

While the technical process of VLMs in attribute inference presents a significant research area~\cite{staab2023beyond,tomekcce2024private}, a critical yet often underexplored dimension is the user's perspective on these capabilities~\cite{zhang2025through}. Understanding how users perceive the possibility of their sensitive attributes being inferred by AI, their comfort levels with such practices, their privacy concerns, and their expectations for control and transparency is paramount for fostering user trust and developing responsible AI systems. If users are unaware of or uncomfortable with the inferences being made, it can lead to distrust not only in specific applications but in AI technologies more broadly. Moreover, insights into user perspectives can directly inform the design of more effective privacy-enhancing technologies that resonate with user needs and mental models.

This paper aims to bridge the existing gap by systematically investigating users' perspectives, concerns, and expectations regarding the inference of their sensitive attributes by VLMs from their visual data, through exploring the following research questions (RQs):

RQ1: How do users perceive the capabilities of VLMs to infer their sensitive attributes from visual data and what are the associated privacy risks they identify?

RQ2: What strategies do users employ to manage these perceived risks?
% RQ2: What strategies do users employ to manage these perceived risks, and what are their expectations from platforms and developers for better privacy protection?

RQ3: What are users' challenges to manage their privacy risks of VLM inferences on their videos, and what's their trade-offs and expectations?
% RQ3: How do users describe balancing the utility offered by VLM-powered applications with their concerns about privacy regarding sensitive attribute inference?

To address these research questions, as a preliminary step, we conducted a semi-structured interview on Chinese users (N=17) to gather insights into their perspectives and experiences regarding AI-driven attribute inference from videos updated to social media. Towards RQ1, our findings indicate that users perceive VLMs as capable of inferring a wide array of sensitive attributes with often unsettling accuracy, including geographical locations and contextual activities, demographic profiles like age and gender, and even socioeconomic status. These perceptions are coupled with significant privacy concerns, such as fears of unauthorized identification and doxxing, misuse of personal information, pervasive surveillance, and potential harm from inaccurate algorithmic judgments.

Towards RQ2, users reported adopting diverse mitigation strategies such as proactive content modification (e.g., blurring, mosaics), selective online exposure, and strategic interaction with AI systems. However, they often expressed skepticism regarding the ultimate effectiveness of these personal efforts against advanced AI. Concurrently, users articulate clear expectations for platforms and developers to implement robust, automated privacy safeguards, enhance transparency in data handling, and provide granular user controls.

Towards RQ3, our study reveals that users engage in a complex and individualized trade-off when balancing the utility of VLMs against their privacy concerns. This often involves a hierarchical assessment of information sensitivity, where disclosure for functional benefits is more acceptable if tied to explicit utility and user control, especially over highly sensitive or directly identifiable information. To sum up, the primary contributions of this paper are:

$\bullet$ We qualitatively detail user experiences and perceptions of VLM inference capabilities.

$\bullet$ We document user-identified privacy risks, their mitigation strategies, and perceived effectiveness.

$\bullet$ We articulate user expectations for the design of trustworthy and privacy-respecting VLM systems.

\section{Related Work}

Research on visual privacy issues can generally be categorized into two main approaches: one that focuses on the technology itself such as system-side policy and visual data collection control, and the other that examines users' understanding of AI-driven visual privacy inferences, which is generally limited and inaccurate. 

\subsection{Privacy Risks in Visual Data and AI Systems}

There are various ways to capture visual data, including recording social media videos and wearing smart glasses. These visual data, provided to AI systems or collected by smart devices themselves, often pose privacy risks to the individuals featured in them~\cite{denning2014situ,yao2019privacy,harborth2021evaluating,koelle2015don,shi2021face,bhardwaj2024focus}. For wearers or video shooters, privacy concerns arise largely  from the mismatch between their mental models and the actual stealthy tracking practices of the devices. Harborth et al.~\cite{harborth2021evaluating} highlighted the gap between users' limited understanding and the reality of covert data collection. For bystanders, they also face privacy risks, particularly in public spaces where images or videos captured may contain sensitive information about them. In many if not all instances, bystanders are unaware of being recorded, raising significant privacy issues~\cite{koelle2018your}.

The advent of LLMs and VLMs has further compounded privacy concerns. LLMs have enabled privacy-infringing inferences from previously unseen texts~\cite{bubeck2023sparks}, while VLMs have shown significant improvements on various PAR datasets~\cite{cheng2022simple,castrillon2023evaluation,wang2024pedestrian}, outpacing earlier, non-VLM-based approaches. For instance, Tomekcce et al.~\cite{tomekcce2024private} explored the potential for privacy attribute inference from visual datasets, extending the privacy concerns associated with textual data to the visual modality.

Therefore, privacy concerns are no longer limited to direct data capture but extend into latent inferential risks that users cannot easily perceive or defend against in real-time interactions. Regarding these threats, researchers have proposed various solutions to mitigate the privacy risks. For instance, Jung and Philipose~\cite{jung2014courteous} developed a system that adjusts the privacy settings of the camera based on social context, while PrivacEye~\cite{steil2019privaceye} employed deep learning to disable video recording based on visual cues. Other approaches, such as the ``Life-Tags'' system by Vatavu et al.~\cite{aiordachioae2019life}, abstracted visual data into concepts and tags rather than storing raw images.
% , though this system still faces risks due to reliance on external cloud services for data classification and processing~\cite{aiorduachioae2020aggregating}.

Although many of these solutions aim to mitigate privacy exposure, they often focus on system-side policy controls rather than enhancing user-side awareness. For example, Privacy-Eye~\cite{steil2019privaceye} uses visual scene classification to automatically disable the camera without providing explicit user feedback, limiting user transparency over system actions. In addition, there is still a lack of AI interaction system designs focusing on user operations and feedback, indicating a critical gap in privacy interaction design.

\subsection{User Perspectives on AI-driven Inferences and Visual Privacy}

A significant body of research indicates that users often possess a limited or inaccurate understanding of how their data is used to generate AI-driven inferences, which is quite opaque for normal users without clear reminder. This gap can result in an ``uninformed'' conception of a service's data practices and the mechanisms by which inferences are made~\cite{warshaw2016intuitions,yao2017folk}. Recent work by Zhang et al.~\cite{zhang2024ghost} also finds that users' mental models regarding the inference process, particularly concerning private memories, are notably opaque. This limited awareness is problematic, as it can lead to misinformed online behaviors that not only fail to preserve privacy but may also degrade the quality of personalized services~\cite{kozyreva2021public}.

In such a situation of limited cognition, users may adopt various protective strategies in response to privacy concerns, ranging from actively withholding or strategically sharing information, such as limiting posts on social media~\cite{milne2017information,olson2005study}, to completely avoiding a service, for instance, by declining to purchase a product~\cite{barbosa2020privacy}. In other cases, users may adopt a stance of resignation, feeling powerless and hoping that online services will act in their best interests~\cite{lau2018alexa}. These strategies are often radical and negative for both users and service providers.

Conversely, interventions that increase transparency can significantly alter user perceptions and behaviors. When users can interact with and make sense of their inferred data, for example, through realistic advertising profiles, it can foster a more informed awareness of privacy implications and heighten interest in privacy-protective actions~\cite{barbosa2020privacy,weinshel2019oh}. Due to improved cognition, users can adopt more positive approaches to inducing risks. Studies presenting users with their own inferred data via retrospective surveys and dashboards revealed a tendency to reject the user of sensitive inferences, such as location or demographics, for third-party applications~\cite{wills2011personalized}. Highlighting this potential, Ma et al.~\cite{ma2025raising} demonstrated that an LLM-powered intervention effectively increased participants' awareness of location privacy risks posed by LLMs and encouraged them to value greater control. 
% However, heightened awareness does not uniformly lead to data restriction. Asthana et al.~\cite{asthana2024know} found that after being made aware of data inferences, participants were often willing to consent to the system's use of both inferred and explicitly provided data.

Despite these insights, a limitation persists in much of the existing research. While numerous studies employing interviews, dairy studies and surveys have successfully elucidated how users make sense of inferences in contexts like online behavioral advertising~\cite{eslami2018communicating,rader2020have,yao2017folk}, voice assistants~\cite{lau2018alexa,malkin2022runtime,vimalkumar2021okay}, and chatbots~\cite{folstad2018makes,liu2022effects}, they often stop short of revealing the concrete privacy decisions users make regarding their data in the face of AI inferences. 

\section{Methodology}

We detailed the participant recruitment, interview design and procedure, and the analysis methods sequentially.

\subsection{Participant Recruitment and Demographics}

This IRB-approved study recruited Chinese participants through distributing posters on online platforms. We recruited 17 participants (9 males, 8 females). The demographics of participants were shown in Table~\ref{tbl:demographic}. Each participant was compensated 100 RMB for their time.

\begin{table}[htbp]
    \centering
    \caption{Demographic information.} % Optional: Adds a caption to your table
    \label{tbl:demographic}
   \scalebox{0.9}{\begin{tabularx}{0.5\textwidth}{lXll}
      \toprule
      User ID & Use Case & Demographic & \makecell[l]{Highest \\Education} \\
      \midrule
      P1  & AI editor, Doubao & Student & Bachelor \\
      P2  & VLMs & Student & Bachelor \\
      P3  & ChatGPT, Doubao & Student & Master \\
      P4  & Doubao, kimi, qwen & NA & Bachelor \\
      P5  & \makecell[l]{Doubao, deepseek, \\Wenxin Yiyan, qwen} & Teacher & Bachelor \\
      P6  & \makecell[l]{deepseek, Doubao, \\Wenxin Yiyan, kimi} & Student & Bachelor \\
      P7  & Sora, deepseek, GPT & Student & Master \\
      P8  & GPT4, Claude 2.0 & Student & Bachelor \\
      P9  & gpt & \makecell[l]{Product\\ Manager} & Master \\
      P10 & Doubao & Student & Bachelor \\
      P11 & NA & Worker & \makecell[l]{High School \\or Lower} \\
      P12 & Doubao & Driver & \makecell[l]{High School \\or Lower} \\
      P13 & Deepseek, Doubao & NA &\makecell[l]{ High School \\or Lower} \\
      P14 & NA & Student & Bachelor \\
      P15 & Doubao & Student & Bachelor \\
      P16 & NA & Student & Bachelor \\
      P17 & Deepseek & Student & Bachelor \\
      \bottomrule
    \end{tabularx}}
    \label{tab:user_cases} % Optional: For referencing the table in your text
  \end{table}

\subsection{Interview Design and Procedure}

We focused on identifying users' mental model towards the inference, their perceived privacy risks and harms of the inference, their own inference capabilities and the comparison with LLMs, their mitigation strategies and the effectiveness. The interview was carried out in a semi-structured manner, with all interviews audio-recorded and transcribed. Each interview lasted about 30 minutes on average.

\subsection{Data Analysis}

We adopted thematic analysis~\cite{clarke2014thematic} on the transcribed data and the original recordings. Two researchers first read through the data for several times to get familiar with the data. They then open coded a subset (3 participants), discussed and solved the disagreements, forming the initial codebook. Using this codebook, these two researchers separately coded the rest of the data, reaching an inter-rater reliability of 0.9 using Cohen's Kappa. 

\section{Results}

We presented the results according to the research questions, first analyzing the perceptions of inference capabilities and privacy risks, and then users' mitigation strategies and their effectiveness, and finally the challenges, trade-offs and expectations.

\subsection{RQ1: User Perceptions of VLM Inference Capabilities and Associated Privacy Risks}

\subsubsection{Users' Perception of VLM Inference}

Users' perception of VLM inference are centered around specific inference types and cases, reflecting their mental model of the ubiquitous inference in the real world. % The perceptions of participants were shown in Table~\ref{tbl:capabilities}.

\paragraph{Ubiquitous location and contextual activity inference.} A predominant theme was the VLM's strong capability to infer geographical location and contextual activities, a finding reported by a significant majority of participants (14/17, 82\%). Models achieved this by leveraging distinct visual cues, such as identifying a specific city street from billboards (P1), deducing a tour group's location from local snacks like \textit{``Roujiamo''} (P8), or recognizing a hometown's specific \textit{``blacksmithing flower''} festival and its cultural status from the scenery (P17). This capability often reached a startling level of precision, with a substantial subset of participants (9/17, 52.9\%) reporting instances of such direct location inference. P15 expressed shock when a VLM identified their specific street from a photo of a pigeon, stating, \textit{``I was dumbfounded, felt like a complete privacy leak.''} While some users found this specificity useful for identifying places (P10), the accuracy of the capability was inconsistent. As P12 noted, a VLM could correctly identify a province but fail in the specific city, demonstrating variable reliability: \textit{``the first time it was right, but later it wasn't.''}

\paragraph{Demographic profile inference: age and gender.} A majority of participants (13/17, 76\%) reported that VLMs can infer basic demographic information like age and gender from visual cues. The inference of gender was perceived to be straightforward. P2 suggested, \textit{``if a person appears, it should be very easy to tell,''} while P9 stated that the probability of correctly identifying male or female \textit{``is quite high.''} P8 provided a detailed account of visual features used by models for analysis, like \textit{``wrinkles, crow's feet''} for age and facial structure and presentation: \textit{``males have that jawline, a bit squarer ... female faces are finer ... then there's hairstyle and makeup.''} for gender.
% explaining that models can infer age from markers like \textit{``wrinkles, crow's feet''} and gender from facial structure and presentation: \textit{``males have that jawline, a bit squarer ... female faces are finer ... then there's hairstyle and makeup.''}
% Inferring basic demographic information such as age and gender is another common capability reported by participants (13/17, 76\%). VLMs are commonly perceived to infer demographic attributes such as age and gender from visual, and to a lesser extent, auditory cues. Gender is often considered relatively easy to infer. P2 suggested, \textit{``For example, gender, with current facial recognition ... if a person appears, it should be very easy to tell.''} P9 stated, \textit{``Gender, if analyzed as male or female, the probability of being correct is quite high.''} The methods involve analyzing facial features, hairstyles, attire, and body shape. P8 provided a detailed account: \textit{``For example, older people ... will have wrinkles, crow's feet ... for facial detection. And females and males are different ... males have that jawline, a bit squarer ... female faces are finer ... then there's hairstyle and makeup.''} 

Age inference is also frequent, as discussed by 11/17 participants (64.7\%), though often guessed as a range. P8 noted that VLMs could infer age based on visual cues (\textit{``facial features ... clothing''}) and even speech patterns: \textit{``Young people like us, born in the 00s or 90s, speak faster. And the words used are very trendy... This model will classify us as younger.''}  P16 observed, \textit{``From the background, and looking at the person's general facial features, age can probably be judged.''} P10 humorously noted a behavioral inference of age: \textit{``Riding a bike with one hand while filming, only someone under 25 would do that.''}

\paragraph{Assessing socioeconomic status and lifestyle indicators}

VLMs are perceived to infer socioeconomic status (income, education, occupation) and lifestyle traits (marital status, hobbies), though with variability in accuracy and reliance on indirect cues. 7/17 (41\%) mentioned income inference, 5/17 (29\%) for occupation, and 5/17 (29\%) for education. For socioeconomic attributes such as occupation, education level, income, and personal interests, these are generally considered challenging to infer accurately without explicit cues. Occupation might be guessed from attire or context, as P2 mused: \textit{``Occupation might consider logic ... the clothes worn, the person's actions, or the logical connection of their actions.''} However, P6 found an inferred occupation (``engineer'') inaccurate. P17 suggested that past interactions could inform such inferences: \textit{``If you are a student ... and you've asked it questions about your homework ... the large model might already have an impression of you ... and estimate your education level and identity.''} P2 also theorized that education level could be inferred if \textit{``someone attending a class in a university or displaying class content.''} was showed. For income, which P2 noted is \textit{``quite abstract''}, direct proof like a \textit{``payslip in the video''} would be definitive. P8 suggested income could be inferred from home decor: \textit{``For example, uploading a video about home decoration and renting a house ... it can infer a Nordic furniture style, and then infer the family's income based on the furniture.''} P17 similarly proposed that VLMs could \textit{``analyze my income situation through the items filmed,''} distinguished between \textit{``high-end places''} and \textit{``crowded activities''} to gauge financial status. Lifestyle and interests are inferred from the activities or objects depicted. P5 noted that a video with a cat might lead to inferences about gender (\textit{``young female''}), age and preferences (\textit{``like small animals''}). P8 also mentioned that preferences for \textit{``minimalist or modern style''} versus more traditional items could be indicative of age and lifestyle. Overall, while these attributes are subjects of VLM inference, their accuracy is more questionable and context-dependent, as expressed or implied by 11/17 participants (64.7\%). P7 for example, stated that inferring education requires specific evidence like a \textit{``student ID''}, and location inference needs \textit{``landmarks or road names''} to be accurate, implying difficulty otherwise.

\paragraph{Perceived Sophistication of Inference and Emergent Privacy Concerns}

Users also found the inference mechanisms advanced and unsettling, leading to negative emotions such as fear, shock or violated feelings (5/17, 29\%). Several participants suspected that VLMs might access more data than explicitly provided, or cross-reference information from various sources. P5 described a particularly \textit{``terrifying''} experience: \textit{``Once I only sent my own photo, it directly stated the photo's location ... it started showing my personal information not given in its ``deep thinking'' process ... It displayed my name and shooting location.''} This user also speculated about photo album access: \textit{``maybe it can read my photo album information, where the photos were taken, where my permanent residence is, it might all be read.''} P2 was aware of metadata like GPS in images but believed their VLM (Doubao) did not typically use it for image content analysis. P17 theorized about data aggregation from other platforms: \textit{``it might collect data from Xiaohongshu or Douyin where users are synchronously posting similar videos, and then analyze my video based on that.''} P11 also believed AI uses \textit{``big data comparison to verify a user's age and other basic information.''} The unexpected revelation of personal data, as in P5's case, caused considerable alarm. 
% These concerns about opaque ``deep thinking'' processes, potential access to broader data ecosystems, and the linking of disparate information points were voiced by 4 participants (P2,P5,P11,P17, 23.5\%), highlighting a tension between VLM capability and user comfort.

\subsubsection{Perceived Privacy Risks and Harms}

Based on the thematic analysis of participants' discussions about VLMs, several distinct categories of privacy risks and underlying concerns emerge. They were shown in Table~\ref{tbl:risks}. Users' mental models of VLMs often depict them as powerful data aggregation and inference engines that learn from vast datasets, store uploaded information, and possess capabilities surpassing human understanding or control, leading to these apprehensions. The primary categories of perceived risks identified are: (1) Unauthorized identification, doxxing and associated harassment or safety threats; (2) Misuse and exploitation of personal information for commercial or malicious ends; (3) Pervasive surveillance, algorithmic profiling, and loss of anonymity; (4) Harm from inaccurate inferences and algorithmic misjudgment.

\begin{table}[htbp]
    \centering
    \caption{User perceptions of VLM inference privacy risks.} % Optional: Adds a caption to your table
    \label{tbl:risks}
    \begin{tabularx}{0.5\textwidth}{p{3.5cm}p{4.5cm}}
      \toprule
      VLM privacy risks & Description\\
      \midrule
      Unauthorized identification, doxxing and associated harassment or safety threats & VLMs enable user identification, leading to doxxing, harassment, and real-world safety threats \\
      \hline
      Misuse and exploitation of personal information for
      commercial or malicious ends & Exploitation of inferred data for commercial manipulation and malicious scams \\
      \hline
      Pervasive surveillance, algorithmic profiling, and loss of anonymity  & Pervasive data collection and profiling erodes anonymity by linking digital and real-world identities \\
      \hline
      Harm from inaccurate inferences and algorithmic misjudgment & Inaccurate VLM inferences create false profiles, leading to misrepresentation and flawed social categorization \\
      \bottomrule
    \end{tabularx}
    \label{tab:user_cases} % Optional: For referencing the table in your text
  \end{table}

\paragraph{Unauthorized identification, doxxing, and harassment}

8/17 participants expressed significant concern that VLMs could enable precise user identification and \textit{``doxxing,''} leading to severe real-world consequences. They feared VLMs can meticulously analyze content to \textit{``locate you accurately, and even excavate all your information''} (P3). This capability is perceived as dangerous because it could democratize doxxing. P4 worried if \textit{``AI can infer this information easily, ordinary people can also infer user information, which is very scary.''} The underlying mental model is that VLMs compile and cross-reference data points to make non-obvious connections. P1 provided an example, from \textit{``a family photo, or a graduation photo ... [it] can deduce which university I attended, and by inferring the time, when I graduated.''}

Participants primarily feared the downstream effects of such identification, including online harassment and \textit{``cyberbullying''} (P4). P12 worried about \textit{``social death''} from viral content and the general risk of exposure: \textit{``I am very worried that current online technology is too powerful. I would be worried about my personal information being posted.''} P17 noted this risk can be gendered, fearing that if men \textit{``determine I'm in the same city, they might harass me via private messages.''} Beyond online threats, 3 participants explicitly linked location inference to their physical safety. P6 stated, \textit{``If it can infer a specific Chinese location correctly, for someone with malicious intent, my personal safety is not guaranteed.''} Similarly, after a VLM identified their garden, P15 feared not only theft but also direct \textit{``harassment or stalking.''}

\paragraph{Misuse and exploitation of personal information for commercial or malicious ends}

7 participants feared that entities could misuse their VLM-inferred personal data for exploitative purposes, ranging from manipulative advertising to outright scams. Participants viewed their data as a valuable commodity, with P2 identifying an \textit{``information asymmetry''} where companies can \textit{``sell this information as a resource.''} This fear of commercialization was specific: P9 believed targeted job content serves to \textit{``commercialize later,''} while P16 strongly objected to their information being sold to \textit{``advertisers or telemarketers,''} leading to unwanted \textit{``promotional calls.''}

These concerns escalated to criminal exploitation. P8 expressed a concrete fear that leaked personal details could allow scammers to \textit{``impersonate bank staff ... and take all my money.''} This reflects a deep mistrust of third-party data handling, which P16 found \textit{``quite uncomfortable''} and P9 worried would lead to \textit{``information leakage or being sold.''} P9 contextualized this ultimate risk by referencing \textit{``Snowden's case''} and the potential for \textit{``espionage.''}

\paragraph{Pervasive surveillance, algorithmic profiling, and loss of anonymity}

6 participants (P1,P2,P5,P7,P9,P11) expressed a fundamental anxiety that VLMs erode privacy through perceived surveillance, deep profiling, and indefinite data storage. This anxiety stems from the mental model that \textit{``AI learns through big data, and things sent are stored in a database''} (P1), fueling fears of \textit{``privacy leakage.''} P5 directly questioned the necessity and risk of this data collection: \textit{``Why does it need to collect my information? If someone searches for something related to me, will it directly tell the searcher my information? Isn't that a huge risk for me?''} This concern highlights a core desire to anonymity online, as another participant stated, \textit{``If other people know everything when I go online, then there's no privacy.''}

Worries also target the volume and granularity of the data processed. P7 feared the leakage of \textit{``professional information, or other physical privacy,''} while P9 expressed concern over corporate surveillance via methods like taking \textit{``periodically take screenshots of my computer.''} The VLM's ability to infer non-explicit information, especially location (P11,P13,P15), further amplifies this sense of lost anonymity. P15 explained that an AI’s superior database allows it to \textit{``know which city you are in''} from a picture when a human cannot, enabling services that could also facilitate unwanted tracking.

\paragraph{Harms from inaccurate inferences and algorithmic misjudgment}

% A distinct risk category, noted by P8, involves the potential harm from VLMs making inaccurate inferences or misjudgments, leading to misrepresentation or unwanted social categorization. While much anxiety focuses on accurate inference of sensitive data, P8 provided a unique example: \textit{``If in my video ... there's Batu Caves ... which has some religions like Buddhism, and some Indians, it might identify me as a local Indian who likes Buddhism. This is wrong information.''} The concern here is the VLM creating a false or skewed profile rather than simply data exposure.

% The consequence of such misinterpretation, according to P8 could be receiving unwanted and irrelevant content (\textit{``on social platforms and my recommendations, I might get more local Indians and Buddhists''}) or even having this ``erroneous information'' affect social perceptions or be ``disclosed to others''. This highlights a nuanced understanding of privacy risks, where the harm comes not from the truth being revealed, but from an algorithmic ``falsehood'' being generated and propagated, potentially impacting one's digital identity and experiences.
Participants described a distinct risk: harm caused not by accurate surveillance, but by inaccurate VLM inferences that lead to misrepresentation. P8 provided a specific example, worrying that a video filmed near a landmark could generate a false profile: \textit{``If in my video ... there's Batu Caves ... which has some religions like Buddhism, and some Indians, it might identify me as a local Indian who likes Buddhism. This is wrong information.''}

P8 further explained the consequences of such misjudgment could range from receiving irrelevant, targeted content (\textit{``on social platforms and my recommendations, I might get more local Indians and Buddhists''}) to having this \textit{``erroneous information''} negatively affect social perceptions or be \textit{``disclosed to others.''} This highlights a nuanced concern where the harm stems not from an exposed truth, but from a propagated algorithmic falsehood that misrepresents a user's digital identity.

\subsubsection{Places Where Participants Perceive Privacy Risks}

% Participants perceive privacy risks associated with VLMs across a spectrum of digital activities, particularly when creating, sharing or interacting with multimedia content that VLMs can analyze. These scenarios range from everyday content sharing to direct engagement with AI systems for personalized tasks, and they were shown in Table~\ref{tbl:scenarios} The primary contexts in which users feel these VLM-related privacy risks manifest can be categories into three main themes: the public sharing of everyday life and social activities, the dissemination of content rich in explicit personal and environmental identifiers, and direct engagement with AI systems for personalized analysis or assistance.
Participants perceive VLM-related privacy risks across various digital activities where they create or share multimedia content, as detailed in Table~\ref{tbl:scenarios}. These contexts fall into three primary categories: (1) the public sharing of everyday life and social activities, (2) the dissemination of content rich in personal and environmental identifiers, and (3) direct engagement with AI systems for personalized assistance.

\begin{table}[htbp]
    \centering
    \caption{Scenarios where participants perceive privacy risks.} % Optional: Adds a caption to your table
    \label{tbl:scenarios}
    \begin{tabularx}{0.5\textwidth}{p{3.5cm}p{4.5cm}}
      \toprule
      Scenarios of privacy risks & Description\\
      \midrule
      Public sharing of everyday life and social activities & Inference risks from publicly shared daily life content. \\
      \hline
      Content rich in explicit personal and environmental identifiers & VLM extraction of explicit identifiers from shared content. \\
      \hline
      Direct engagement with AI for personalized tasks or information disclosure & Direct interaction with AI enables unintended user profiling. \\
      \bottomrule
    \end{tabularx}
    \label{tab:user_cases} % Optional: For referencing the table in your text
  \end{table}

\paragraph{Public sharing of everyday life and social activities}

% A prevalent context for perceived VLM-related privacy risks involves the creation and public dissemination of content depicting everyday life and social activities, often through vlogs or social media platforms. 5 participants (P1,P2,P12,P13,P17) highlighted scenarios where causal documentation of daily routines or social gatherings could lead to unintended privacy breaches when processed by VLMs. P1 stated that \textit{``causally filming life vlogs can inadvertently reveal locations, landmarks, road signs, etc., making it easy to infer my place of residence and lifestyle.''} Similarly, P17 described how filming an outing with friends might \textit{``leak our city, our ages, and regional information,''} and even a video taken at a bar could \textit{``directly pinpoint the exact location of the bar I was in.''} The mental model here is that VLMs can extract and synthesize seemingly innocuous details from such videos to build a comprehensive profile or identity specific locations. P12 expressed anxiety about a humorous video with friends posted on Douyin, fearing that if it \textit{``suddenly becomes popular ... I would experience social death,''} because AI might \textit{``push it to everyone,''} making a private moment uncomfortably public. This indicates a concern that VLMs ampligy the reach and analytical depth applied to content shared in semi-public or public digital spaces.
5 participants (P1,P2,P12,P13,P17) identified significant privacy risks in publicly sharing content about their everyday lives and social activities. They worried that casual documentation could lead to unintended data breaches. P1 noted that by \textit{``causally filming life vlogs,''} a VLM could easily infer their \textit{``place of residence and lifestyle''} from landmarks and signs. Similarly, P17 described how a video of a social outing might \textit{``leak our city, our ages, and regional information,''} or even \textit{``directly pinpoint the exact location of the bar I was in.''}

The underlying fear is that VLMs extract and synthesize these seemingly innocuous details to build comprehensive user profiles. This can lead to severe social consequences, as P12 feared \textit{``social death''} if a casual video posted on Douyin \textit{``suddenly becomes popular''} because the AI might \textit{``push it to everyone.''} This illustrates a core concern that VLMs dangerously amplify both the audience and the analytical depth applied to content shared in public digital spaces.

\paragraph{Content rich in explicit personal and environmental identifiers}

% Users feel a heightened sense of privacy risk when the content they share or interact with is rich in explicit personal or environmental identifiers, which VLMs are perceived to be adept at processing. This was a concern for 9 participants (P1,P3,P4,P5,P6,P8,P10,P11,P13,P16). Scenarios include videos or images containing recognizable faces (P3,P6,P16), specific locations like homes, schools, or unique landmarks (P1,P4,P5,P8,P11,P13,P16), and textual information such as ID cards, tickets, on-screen text, or even street signs (P3,P4,P10). P4 emphaszied that \textit{``if the image contains specific text, it's easy to expose information ... AI's ability to capture text from videos is not bad''}, and \textit{``surrounding buildings, even if mosaic-ed, can often be inferred, and AI recognition is more accurate.''} P5 shared a direct experience: \textit{``I uploaded my life photos ... it could directly infer my age ... and by inferring the appearance of trees or the campus outside, it could deduce where my neighborhood is, exposing my address.''} The underlying fear is that VLMs can easily extract these explicit identifiers, link them, and create highly detailed and potentially sensitive profiles. P10 listed examples such as \textit{``people photographing their first flight, train journey, friends' faces, or important information on tickets.''} P16 also noted that filming content that includes \textit{``a university platform''} or features \textit{``personal appearance''} can lead to easy inference of student status or other details.

9 participants expressed heightened privacy risk when their content contains explicit personal or environmental identifiers that VLMs adeptly process. They identified risks in sharing recognizable faces, specific locations like homes or schools, and textual information on tickets or signs. P4 emphasized the VLM's superior capability, stating that \textit{``if the image contains specific text, it's easy to expose information,''} and that even obscured \textit{``surrounding buildings ... can often be inferred.''}

P5 shared a direct experience where a VLM processed their life photos to \textit{``directly infer my age ... and by inferring the appearance of trees or the campus outside, it could deduce where my neighborhood is, exposing my address.''} This reflects a core fear that VLMs can easily extract and link these explicit identifiers to build detailed profiles. Participants listed many examples of such risky content, including photos of a \textit{``first flight, train journey, friends' faces''} (P10) or videos featuring a \textit{``university platform''} (P16).

\paragraph{Direct engagement with AI for personalized tasks or information disclosure}

% Privacy risks are significant when users directly engage with VLMs for personalized tasks, especially when this involves uploading sensitive personal content or providing PII, as explicitly noted by 4 participants (P3,P5,P6,P7). P5 recounted uploading \textit{``personal photos''} leading to age and location inference, and worried that providing enough information during interactions could even lead to the inference of an \textit{``ID card number.''} P6 described a high-risk scenario as \textit{``telling it my identity, student ID, or even photos of myself when asking questions based on my personal situation.''} Similarly, P7 expressed concern that \textit{``if I ask the large model some quetions related to my work or study, it might infer that I am a student, and if I reveal more information, it will know where I am.''} The mental model operating here is that information directly fed to the VLM is not only processed for the immedaite task but also potentially stored, learned from, and used to build a comprehensive profile of the user (P1: \textit{``AI learns through big data, and things sent are stored in a database''}). This direct provision of, or querying about personal data makes the inference capabilities of VLMs feel particularly intrusive and the potential for misuse more direct. P3 noted that risks are heightened \textit{``when I ask it some specific questions about this video,''} suggesting that the act of inquiry itself can guide the VLM towards sensitive disclosures.

4 participants (P3,P5,P6,P7) identified significant privacy risks when directly engaging VLMs with sensitive personal data for personalized tasks. They worried that providing such information could lead to unexpected and deep inferences. P5 recounted how uploading \textit{``personal photos''} could lead to the inference of an \textit{``ID card number,''} while P6 described \textit{``telling it my identity, student ID, or even photos of myself''} as a high-risk scenario. Similarly, P7 feared that asking \textit{``questions related to my work or study''} would allow a VLM to infer their location.

This anxiety is rooted in the mental model that directly provided data is not just used for the immediate task but is permanently stored and learned from. As P1 stated, \textit{``AI learns through big data, and things sent are stored in a database.''} This makes the VLM's inference capabilities feel more direct and intrusive. P3 added a further nuance, noting that risk is heightened even when merely inquiring about data, as \textit{``when I ask it some specific questions about this video,''} the act of inquiry itself can guide the VLM toward sensitive disclosures.

\subsection{RQ2: User Mitigation Strategies and Their Effectiveness}

% Participants employed various strategies to mitigate the perceived privacy risks associated with VLMs processing their multimedia content. These measures range from direct content alteration and controlled sharing practices to mindful interaction with AI systems and reliance on external safeguards, reflecting a conscious effort to reclaim some control over their digital footprint. These measures and their effectiveness in participants eyes were shown in Table~\ref{tbl:strategies}.  The primary categories of these avoidance strategies include: proactive content modification and obfuscation; selective disclosure and controlled online exposurel; strategic interaction and data minimization when engaging with AI systems; a combination of reliance on external controls and personal vigilance in content management; and an acknowledgement of the inherent limitations in fully mitigating these risks.
To mitigate the perceived privacy risks of VLMs processing their multimedia content, participants employed various strategies, as detailed in Table~\ref{tbl:strategies}. These strategies fall into five primary categories: (1) proactive content modification and obfuscation, (2) selective disclosure and controlled online exposure, (3) strategic interaction with and data minimization for AI systems, (4) a combination of reliance on external controls and personal vigilance, and (5) an acknowledgement of the inherent limitations in fully mitigating these risks.

\begin{table}[htbp]
    \centering
    \caption{Mitigation strategies and their effectiveness.} % Optional: Adds a caption to your table
    \label{tbl:strategies}
    \begin{tabularx}{0.5\textwidth}{p{3.5cm}p{4.5cm}}
      \toprule
      Mitigation strategies & Effectiveness \\
      \midrule
      Proactive content modification and obfuscation & Directly sanitizes data, disrupting VLM data extraction \\
      \hline
      Selective disclosure and controlled online exposure & Fundamentally limits data available for VLM analysis \\
      \hline
      Strategic interaction and data minimization with AI systems  & Offers partial control during necessary AI interactions \\
      \hline
      Reliance on external controls and vigilant personal review & Strong protection through meticulous self-monitoring and review \\
      % All strategies & VLMs possess almost omniscient analytical power, making strategies less useful; Adoption of preserving measures provides psychological comfort and alleviate anxiety \\
      \bottomrule
    \end{tabularx}
    \label{tab:user_cases} % Optional: For referencing the table in your text
  \end{table}

\subsubsection{Proactive content modification and obfuscation}

% A frequently cited strategy, mentioned by 7/17 participants (P1,P2,P3,P4,P10,P13,P15), involves directly modifying or obfuscating multimedia content before uploading or sharing it, specifically to prevent VLMs from extracting sensitive information. The most common technique is the application of mosaics or blurring. P1 stated they would \textit{``apply mosaics to places or road signs that I clearly don't want to reveal,''} and P13 described blurring and cropping images for an AI medical consultation to \textit{``only leave the part to be checked, equivalent to applying a mosaic.''} P2 considered mosaics the \textit{``simplest form of desensitization''} for sensitive information like ID numbers or faces, believing it causes \textit{``permanent destruction to the image layer.''} Other modifications include P3 removing specific audio segments if they contain sensitive information, or P2 suggesting the removal of image metadata (location, time, camera model) using software, a more technical form of desensitization. This theme underscores a hands-on approach by users to sanitize their data before it can be processed by potentially intrusive VLM analyses.
7/17 participants (P1,P2,P3,P4,P10,P13,P15) described directly modifying or obfuscating their multimedia content to prevent VLMs from extracting sensitive information. The most common technique was applying mosaics or blurring. P1 would \textit{``apply mosaics to places or road signs that I clearly don't want to reveal,''} while P13 blurred an image for a medical consultation to \textit{``only leave the part to be checked.''} P2 considered this the \textit{``simplest form of desensitization,''} believing it causes \textit{``permanent destruction to the image layer.''} Other modifications included removing sensitive audio segments (P3) or stripping image metadata, which P2 suggested as a technical form of desensitization. These actions demonstrate a hands-on approach by users to sanitize their data before it undergoes potentially intrusive VLM analysis.

\subsubsection{Selective disclosure and controlled online exposure}

% 10/17 participants (P1,P2,P4,P5,P6,P7,P9,P11,P15,P16) adopt strategies of selective disclosure, consciously deciding what content is too sensitive to record or upload in the first place, thereby limiting the data available to VLMs. P2 mentioned that the \textit{``risk of privacy infringement is an important reason preventing me from doing self-media.''} P5 stated, \textit{``I try to prevent it from reading my information by ... avoiding filming my life scenes, or things that might expose my address, occupation, or work environment.''} This selective creation extends to choosing audiences. 4 participants (P4,P8,P9,P12) described methods like using privacy settings on social media platforms (P4), setting video visibility to \textit{``close friends only''} (P8), sharing \textit{``only in smaller circles''} (P9), or making content private/followers-only and using niche tags to \textit{``reduce traffic to the minimum''} (P12). This reflects an understanding that limiting exposure can reduce the surface area for VLM analysis and subsequent privacy risks. P16 explicitly mentioned trying to \textit{``avoid personal facial information and landmark objects appearing [in videos], and minimize the time they appear on screen.''}
10/17 participants practice selective disclosure, consciously deciding what content is too sensitive to record or upload to limit the data available to VLMs. This can be an absolute refusal, as P2 noted the \textit{``risk of privacy infringement is an important reason preventing me from doing self-media.''} More commonly, participants practice selective creation. P5 tries to \textit{``avoid filming my life scenes, or things that might expose my address,''} and P16 attempts to \textit{``avoid personal facial information and landmark objects appearing [in videos].''}

This strategy extends to controlling the audience. 4 participants use platform privacy settings to share content with \textit{``close friends only''} (P8), \textit{``only in smaller circles''} (P9), or use niche tags to \textit{``reduce traffic to the minimum''} (P12). These practices demonstrate an understanding that limiting exposure reduces the surface area for VLM analysis and its subsequent privacy risks.

\subsubsection{Strategic interaction and data minimization with AI systems}

% When direct interaction with VLMs is necessary, 5 participants (P3,P5,P6,P7,P9) described strategies aimed at minimizing the data provided to or inferable by these AI systems. This includes being cautious about the nature of queries and the information explicitly shared. P3 would \textit{``choose to ask more general questions''} if needing to submit a video but wishing to limit information leakage. P6 advised to \textit{``not give the large model too much specific information''} and suggested instructing the VLM to \textit{``anonymize all privacy-implicating information with symbols''} in its output. P5 detailed a multi-pronged approach: \textit{``manage camera permissions to only be active when I use it ... if there's a situation where it reads my information, I tend to clear big data history, reduce interaction with it, and manage permissions for big data access.''} P7 also emphasized trying to \textit{``minimize giving it useful information.''} P9 acknowledged the difficulty of complete avoidance with persistent use of a single AI account but suggested that \textit{``informaiton hiding''} or pre-setting contexts with the AI could be partial measures. Through this measure, users develop a cautious engagement model with VLMs, treating them as entities that require careful information sharing.
5 participants (P3,P5,P6,P7,P9) described strategies for minimizing data disclosure during necessary interactions with VLMs. To control their input, they limit the specificity of their queries. P3 would \textit{``choose to ask more general questions,''} and P7 would try to \textit{``minimize giving it useful information.''} P6 advised to \textit{``not give the large model too much specific information''} and even suggested instructing the VLM to \textit{``anonymize all privacy-implicating information with symbols''} in its output.

Participants also use technical controls. P5 detailed a multi-pronged approach including managing \textit{``camera permissions,''} clearing \textit{``big data history,''} and reducing \textit{``interaction with it.''} While acknowledging the difficulty of complete avoidance on a single account, P9 suggested partial measures like \textit{``information hiding.''} These strategies demonstrate a cautious engagement model, where users treat VLMs as entities requiring careful information sharing.

\subsubsection{Reliance on external controls and vigilant personal review}

A notable group of participants indicated reliance on external controls, such as platform policies or emphasized personal vigilance through careful content review. Among them, 2 participants (P6,P7) considered the privacy policies or statements of VLM providers. P7 stated, \textit{``I tend to choose models that [their company] first declares they will not steal my privacy. If they don't have such a declaration ... I will not transmit information to such models.''} This shows a desire for VLM developers to take responsibility for user privacy. Complementing this, 3 participants (P8,P14,P17) highlighted the importance of personal review and post-upload control. P8 would \textit{``first check the uploaded content, then apply mosaics to things I feel cannot be shown, and then upload it.''} P14, if accidental exposure was found, would \textit{``immediately delete the work.''} P17 described a meticulous approach: \textit{``I might review my videos multiple times to see if any points that could leak information... were accidentally filmed... and I might be less active in uploading to media.''} This suggests that some users combine cautious selection of tools with diligent self-monitoring.

\subsection{Acknowledged limitations, risk acceptance, and perceived high mitigation costs}

Participants acknowledged the inherent difficulty and high cost of fully mitigating VLM-related privacy risks, often leading to a pragmatic acceptance of exposure. P9 explained this trade-off, noting that the \textit{``cost of hiding information is very high,''} especially for video, so \textit{``many people cede a part of their privacy rights.''}

This difficulty can result in a passive stance, a sentiment shared by two participants (P2,P9). For instance, despite knowing about desensitization techniques, P2 admitted to \textit{``currently not taking any measures to avoid risks''} for personal content, citing a \textit{``not very strong privacy awareness.''} This suggests that for some users, the high effort required for protection, combined with the fact that \textit{``short-term risks from privacy leakage are not visible,''} leads to a conscious acceptance of a certain level of privacy risk.

% Participants acknowledged the inherent difficulties in completely avoiding VLM-related privacy risks, the high cost associated with thorough mitigation, and expressed a high degree of risk acceptance. P9 provided a detailed perspective: \textit{``Many people cede a part of their privacy rights because it's very hard to 100\% eliminate this concern... the cost of hiding information is very high. If you upload a picture, you can apply a mosaic. If you upload a video of several minutes, it's very difficult to simply and specifically hide various kinds of information.''} This highlights a pragmatic, if somewhat resigned, view that the convenience of necessity of using VLM-integrated platforms comes at a privacy cost that is hard to fully negate. P2, while aware of desensitization techniques, admitted to \textit{``currently not taking any measures to avoid risks''} for personal photos and videos, perceiving them as \textit{``not very sensitive''} at present and having a \textit{``not very strong privacy awareness''} for personal data, though acknowledging future concerns. This sentiment, shared by 2 participants (P2,P9), suggests that for some, the perceived immediacy of risk or the effort required for comprehensive protection leads to a more passive stance or an acceptance of a certain level of exposure, especially when \textit{``short-term risks from privacy leakage are not visible.''} 

\subsection{Effectiveness of Mitigation}

Participants held varied and nuanced views on the effectiveness of their strategies to mitigate VLM-related privacy risks. While some believed their measures offered tangible protection against casual observation or simple inference, a significant skepticism remained about their robustness against advanced AI or determined actors. Overall, participants viewed any effectiveness as conditional and partial, sometimes valuing their strategies more for psychological comfort than for providing foolproof security.

% Participant perceptions regarding the effectiveness of measures taken to mitigate VLMs-related privacy risks are notably varied and nuanced. While some users believe their strategies offer a tangible degree of protection, particularly against casual observation or less sophisticated inference attempts, a significant undercurrent of skepticism prevails regarding the robustness of these measures against advanced AI capabilities and determined actors. The overall sentiment suggests that effectiveness is often conditional, partial, and sometimes valued more for psychological comfort than for providing foolproof security.

\subsubsection{Conditional and partial efficacy of common mitigation techniques}

Many participants believe common strategies like content modification and selective sharing offer a conditional or partial degree of effectiveness. 4 participants (P1,P4,P10,P16) viewed content obfuscation, primarily mosaics, as a key defense. They opined that \textit{``AI is not yet strong enough to directly see through mosaics''} (P1) and that these methods have \textit{``some effect''} (P4), with P10 estimating them to be \textit{``effective about 80\% of the time.''} However, they acknowledged clear limitations. Mosaics become \textit{``merely symbolic''} for highly recognizable landmarks (P10), and P2 warned that from many desensitized images, \textit{``some information can be inferred because omissions are very common.''}

Selective sharing was also seen as a valuable control; P3 felt that with certain measures, \textit{``at least strangers won't know,''} and P6 trusted their methods were effective with \textit{``reputable companies.''} P8 highlighted a more advanced tactic of using obfuscation to induce misidentification, noting a VLM might misidentify a landmark, and this \textit{``wrong information, in turn, protected my privacy.''} Ultimately, some participants recognized absolute data minimization as the most reliable strategy, as P15 stated simply, \textit{``not filming is very effective.''}

\subsubsection{Skepticism regarding efficacy against advanced AI and determined adversaries}

Despite the perceived benefits of countermeasures, a strong skepticism emerged, with 4 participants (P9,P12,P16,P17) doubting the ultimate effectiveness of their mitigation strategies against sophisticated VLMs. This skepticism is rooted in a belief that the VLM's analytical power, derived from \textit{``big data,''} is overwhelming. P12 was particularly pessimistic, arguing that because \textit{``current technology is very powerful,''} any measure short of total abstinence from posting is \textit{``useless.''} P16 echoed this, feeling that \textit{``effectiveness is not that great, big data is so powerful, it feels like it can analyze all my info at once.''} Similarly, P17 acknowledged that while their efforts might have \textit{``some effect... perhaps not that much''} because the model's analytical ability \textit{``is still very strong.''}

% Despite the perceived benefits of some measures, a strong theme of skepticism emerges, with 4 participants (P9, P12, P16, P17) expressing doubts about the ultimate effectiveness of their efforts against sophisticated VLMs or determined individuals. P12 was particularly pessimistic, stating, \textit{``I feel it's useless because current technology is very powerful... AI can figure it out from just a tiny bit of an architectural feature in a photo of the sky... unless you don't post at all, I feel other methods are useless.''} This reflects a mental model where VLMs possess almost omniscient analytical power derived from ``big data.'' P16 echoed this, feeling that \textit{``effectiveness is not that great, big data is so powerful, it feels like it can analyze all my info at once,''} and P17 similarly acknowledged that while measures might have \textit{``some effect... perhaps not that much because... the model's ability to analyze information is still very strong.''}

% This skepticism is also fueled by personal experiences where VLMs circumvented attempts at control. P5, despite deleting history and limiting interaction, found a VLM still accessed location and music app information, concluding the measures \textit{``didn't have much use for me, but it made me feel my privacy was leaked.''} P9 acknowledged that while measures are \textit{``relatively effective... if someone really wants to investigate you, they can also use illegal means,''} implying that technical defenses have limits against determined human adversaries, with or without AI.

\subsubsection{The role of avoidance measures in providing psychological comfort}

For some participants (2/17, P11, P13), the adoption of privacy-preserving measures serves an important function in providing psychological comfort, even if the actual security benefits are uncertain or perceived as limited. P11 admitted that complete avoidance is \textit{``impossible... but relatively speaking, it gives oneself psychological comfort.''} P13 echoed this sentiment: \textit{``I don't know if it's reliable, but psychologically it feels effective.''} This suggests that engaging in these practices, such as applying mosaics or being cautious with uploads, can create a sense of agency and control in an environment where users often feel their data is vulnerable. The act of taking precautions, regardless of its ultimate impenetrability, can alleviate anxiety and allow users to continue participating in digital spaces, albeit with a cautious mindset. This psychological dimension is an important aspect of user behavior, indicating that the perceived, rather than solely the actual, effectiveness of privacy measures plays a role in their adoption. 

\subsection{RQ3: Challenges, Trade-offs and Expectations}

\subsubsection{Perceived Trade-offs and the High Cost of Complete Privacy}

Users also highlighted the trade-offs between privacy, usability and the effort required for robust protection. While not posting content at all is recognized as highly effective (P15: \textit{``not filming is very effective''}), it fundamentally curtails participation in digital content creation, a trade-off not all users are willing to make. P9 articulated the high cost of comprehensive mitigation: \textit{``The cost of hiding information is very high. If you upload a picture, you can apply a mosaic. If you upload a video of several minutes, it's very difficult to simply and specifically hide various kinds of information.''} This implies that users often engage in a practical cost-benefit analysis, where the effort of meticulous desensitization for every piece of content might outweigh the perceived immediate risk, especially as P2 noted, when \textit{``omissions are very common''} even with desensitization efforts. This can lead to a degree of risk acceptance, as noted by P9, \textit{``many people cede a part of their privacy rights because it's very hard to 100\% eliminate this concern,''} and they may \textit{``default to [accepting] there's a risk, and it doesn't matter if there's a risk.''} This underscores the challenge users face in balancing their desire for privacy with the content sharing aim with pervasive VLMs.

\subsubsection{Expectations of Platforms, Stakeholders and Supervisors}

Participants envision a multi-stakeholder approach to fortifying privacy in the age of advanced VLMs, expressing expectations for proactive measures from platforms and developers, enhanced user control and transparency, strong regulatory frameworks, and a continued evolution of user vigilance. The expectations were shown in Table~\ref{tbl:expectations}. Their future outlook emphasizes a shared responsibility model where technological solutions, ethical design principles, clear governance, and informed user practices converge to mitigate the risks associated with sophisticated AI-driven content analysis.

\begin{table}[htbp]
    \centering
    \caption{An overview of user expectations for privacy protection} % Optional: Adds a caption to your table
    \label{tbl:expectations}
    \begin{tabularx}{0.5\textwidth}{p{4cm}p{4cm}}
      \toprule
      Expectations & Description \\
      \midrule
      Proactive platform-integrated privacy safeguards and automated moderation & Allow AI identify and inform users which information constitutes privacy violations\\
      \hline
      AI-assisted content modification tools & Provide tools to allow users edit parts of the image \\
      \hline
      Enhanced transparency, user control, and granular consent mechanisms  & Let users explicitly know the data handling practices and offer more granular control \\
      \hline
      Strengthened developer responsibility and regulatory oversight & Privacy-by-design principles to be embedded by developers; Government oversight and legal monitoring  \\
      \hline
      Empowered and cautious user practices &  Users take a heightened sense of awareness and more deliberate decision-making regarding content creation and sharing \\
      \bottomrule
    \end{tabularx}
    \label{tab:user_cases} % Optional: For referencing the table in your text
  \end{table}

\paragraph{Proactive platform-integrated privacy safeguards and automated moderation}

A dominant expectation, voiced by a significant majority of participants (10/17) is for VLM-integrated platforms and editing software to incorporate proactive and automated privacy-enhancing features. Users anticipate that AI itself should be leveraged to protect privacy. P1 suggested platforms should \textit{``edit AI's own content, allowing AI to identify which information constitutes user privacy or violations,''} and then offer users options to \textit{``replace, hide, or delete''} problematic sections. This includes automated detection and flagging of sensitive information. P8 envisioned a system where the VLM \textit{``detects and highlights potential leaks (e.g., logos, location, faces), making it obvious to me where information might be exposed.''} P7 suggested VLMs could \textit{``automatically help me mosaic or otherwise prevent access to''} inadvertently captured sensitive details like ID cards.

Furthermore, there's a desire for AI-assisted content modification tools. P4 hoped editing software like Jianying, which already incorporates AI functions, could \textit{``develop reminder functions''} that \textit{``automatically detect aspects that can help privacy, reminding users to apply mosaics or edit parts of the footage.''} P5 suggested platforms could automatically \textit{``apply mosaics to areas imposing risks ... or tell users which parts that might expose privacy should be cropped.''} P8 also proposed features like \textit{``background blur, just showing my face,''} allowing users to \textit{``decide whether to show a face or an outdoor landmark.''} The underlying mental model is that since VLMs are powerful at analysis, they should also be powerful at automated safeguarding, shifting some of the burden from the user to the technology provider.

\paragraph{Enhanced transparency, user control, and granular consent mechanisms}

Complementing automated safeguards, participants strongly desire increased transparency regarding data handling practices and granular control over their personal information, a sentiment expressed by 5 participants. P5 articulated a need for applications to \textit{``clearly tell me they will not read my information, and what penalties they might face if they do,''} advocating for \textit{``clear terms ... letting users explicitly know.''} This extends to permission management, where users are informed \textit{``what permissions are needed, what privacy they involve, whether it's an acceptable range, whether they can be turned off, and if it will affect usage''} (P5). P2, while skeptical around the efficacy of lengthy user agreements that \textit{``99\% of users won't read,''} suggested that ``reminder boxes'' coupled with options like \textit{``one-click desensitization before uploading''} could be effective.

A key aspect of control is data deletion and purpose limitation. P3 hoped for models to \textit{``delete their memory, for example, delete user data''} after processing. P6 wished for an option where \textit{``after generation is complete, a prompt appears asking if the large model should automatically delete this data ... so the data I generate can only be used by myself and is deleted after use.''} P12 envisioned an ``incognito mode'' for AI apps like Doubao, and settings to ensure \textit{``what I send is not authorized for others or for big data, making it completely private.''} These expectations reflect a user desire to not only be informed but also to actively manage the lifecycle and accessibility of their data within VLM systems.

\paragraph{Strengthened developer responsibility and regulatory oversight}

Participants also look towards developers (the ``business-end'') and regulatory bodies to establish a safer VLM ecosystem, a view supported by 4 participants (P2,P9,P14,P17). There's a strong sentiment that the onus of privacy protection should not solely rest on users. P2 asserted that risk mitigation should depend \textit{``more not on customer-end users figuring it out themselves, but on how the business-end ... can avoid company ethics, safety, and privacy risks.''} This implies a call for privacy-by-design principles to be embedded by developers. P9 detailed technical expectations such as anonymization for sensitive information, end-to-end encryption for data transmission, and clarity on data storage practices (local vs. cloud, and potential sharing with other entities or across borders, citing \textit{``political risks''} and data sovereignty concerns like \textit{``data placed in Ireland or Singapore''}).

The role of government and regulatory oversight is also highlighted. P2 stated, \textit{``the government should do more risk screening and control.''} P14, suggesting AI algorithms for risk detection on social platforms, implied a need for standards or enforcement: \textit{``if platforms can detect that you have some risks when you click publish, and a warning box pops up, I think this might be a better way.''} P17 acknowledged that VLM data use \textit{``will also be subject to national supervision and legal monitoring,''} expressing a hope that this oversight ensures platforms \textit{``rigorously review uploads for accidental PII ... and provide feedback to the user.''} This indicates an expectation that external authorities will enforce accountability and baseline privacy standards for VLM operations.

\paragraph{Empowered and cautious user practices}

While emphasizing the responsibilities of platforms and regulators, 5 participants also foresee a future where users themselves adopt empowered and cautious practices. This involves a heightened sense of awareness and more deliberate decision-making regarding content creation and sharing. P10 suggested temporal or geographical displacement of posts (\textit{``If I went to Shanghai today, I'll post about it a few days later ... from a different region''}) or even altering self-presentation (\textit{``dress to look poorer''}). P11 indicated a willingness to \textit{``modify personal information before publishing again''} or \textit{``no longer use AI to make videos in that area''} if risks are identified.

P17 described a future approach involving increased personal diligence: \textit{``I might pay more attention to my uploading behavior... review my videos multiple times... to see if any points that could leak information ... and I might be less active in uploading to media... [I will] control the desire to share and not upload all my daily life to social media.''} This suggests a move towards more cautious participation, where users actively curate their online persona and manage their ``sharing desire'' in response to the capabilities of VLMs. 

\subsubsection{Privacy-utility balancing perspectives}

% Participants engage in a nuanced and often highly individualized calculus when balancing the perceived utility of VLM functionalities against their privacy concerns. This balancing act is not uniform; it varies significantly based on the type of information in question, the specific context of VLM use, and individual user thresholds for privacy. The detailed accounts from participants P15, P16, and P17 in this specific corpus illuminate key aspects of this privacy-utility trade-off, revealing distinct user expectations and delineations for acceptable data use by VLMs.
Participants engage in a nuanced and often highly individualized calculus when balancing the perceived utility of VLM functionalities against their privacy concerns. This balancing act is not uniform. It varies significantly based on the type of information in question, the specific context of VLM use, and individual user thresholds for privacy. 

\paragraph{Hierarchical sensitivity of information and utility-driven disclosure}

A primary factor governing the privacy-utility balance is the perceived sensitivity of the information, with users demonstrating a tiered approach to what they are willing to share. P16 and P17 clearly distinguish between general categorizations or functional data and core personally identifiable or highly sensitive information. P16, for instance, is amenable to VLMs using \textit{``broad category''} information, such as identifying them as a \textit{``student or a running enthusiast,''} to receive ``helpful'' personalized recommendations for \textit{``student supplies''} or \textit{``products they like.''} However, this acceptance sharply contrasts with a strong objection to the disclosure of data that enables direct, unsolicited contact: \textit{``I don't like it when they use my phone number or email for promotional marketing.''} Similarly, P17 is \textit{``willing to tell''} a shopping app their \textit{``size, height, and weight''} for clothes selection, or a news app their preferences for content, as this data directly serves the app's explicit function. This willingness is strictly conditional on utility. P15 suggests that any information obscured (e.g., via mosaics) is \textit{``always unimportant,''} implying a binary view where private information should not be compromised for utility, stating they would \textit{``never trade-off privacy to post something.''} This highlights that while some users engage in a granular cost-benefit analysis per information type, others adopt a blanket protective posture.

\section{General Discussions}

% The study illustrated the significant disconnect between the rapid advancement of VLMs and the privacy perceptions and expectations of the users whose data fuels these systems. The findings reveal a populace that is simultaneously impressed and unsettled by the inferential power of VLMs, struggling to reconcile the utility of AI-powered services with a growing sense of vulnerability. Our discussion will delve into the nuanced psychological and socio-technical implications of these findings, exploring the efficacy and burden of user-led mitigation, the critical need for paradigm shift in transparency and control, and the future trajectory of privacy in an era of pervasive AI analysis.
% % The study illustrated the significant disconnect between the rapid advancement of VLMs, and the privacy perceptions of the users whose data fuels these systems. The findings reveal users' struggling to reconcile the utility of AI-powered services with a growing sense of vulnerability. Our discussion will delve into the nuanced psychological and socio-technical implications of these findings, exploring the efficacy and burden of user-led mitigation, the critical need for paradigm shift in transparency and control, and the future trajectory of privacy in an era of pervasive AI analysis.
% This study highlights a significant disconnect between the rapid advancement of VLMs and the privacy perceptions of users. The discussion focuses on the socio-technical implications of these findings, focusing on the efficacy and burden of mitigation strategies.

\subsection{The Emergence and Prevalence of Inference Risks}

Early works primarily focused on privacy risks unknown to users, such as location information hidden within photo metadata~\cite{henne2013awareness}. There are also many works on social media which focus on other people's privacy~\cite{castrillon2023evaluation,cheng2022simple} (e.g., bystanders~\cite{koelle2018your}), the potential privacy disclosure~\cite{shi2021face} and surveillance~\cite{koelle2015don}. Different from images~\cite{ma2025raising}, videos brought the additional dimension of temporality, which might bring additional privacy risks from inferring these videos. Regarding inferential privacy risks, while users gain some awareness after being explicitly informed, they often struggle to detect these risks independently. Such risks are not only prevalent in widespread social media platforms~\cite{choi2023privacy,ma2025raising} but can also raise during task completion with smart wearable glasses~\cite{cai2025aiget} and on numerous self-logging devices~\cite{barker2024flexible}, potentially leading to diverse threats. Our risks of inference privacy primarily targets the owner itself, as the inferences, and the resulted user profiles are usually interconnected with the social media account~\cite{kozyreva2021public}, rather than on the bystanders shot, and we acknowledged that future work could explore the potential threats around inference even on a bystander recorded in the videos, especially under VLM-induced data aggregation from different sources~\cite{wang2024pedestrian}, which could potentially profile the bystander.

Our work focused on the inference problems of videos, which service providers could use the temporal, and even causal relationships for inference. As evidenced by users, the videos uploaded on the social media are often edited, which means attackers may aggregate videos from different perspectives for inferring users' attributes~\cite{anderson2025making}. This method, similar to self-consistency~\cite{wangself}, may results in stronger inference capability, posing greater harms than inferring from pictures~\cite{ma2025raising}, warranting further attention.

Therefore, we advocate for improved privacy management methods. This call is partly driven by the inadequacy of current consent mechanisms~\cite{asthana2024know}. These mechanisms fail to inform users about inferential risks, much less enable granular control over such privacy-sensitive data. We propose exploring automated, content-aware fine-grained control methods~\cite{venugopalan2024aragorn} and leveraging them to address inferential privacy risks~\cite{zhang2024ghost}. For instance, this could involve performing inference using on-device models, presenting the inferred information to users, and allowing them to edit it via an interface.

% Earlier works primarily focused on privacy risks unknown to users, such as location information hidden within photo metadata~\cite{henne2013awareness}. Different from images~\cite{ma2025raising}, videos brought the additional dimension of temporality, which might bring additional privacy risks from inferring these videos. Regarding inferential privacy risks, while users gain some awareness after being explicitly informed, they often struggle to detect these risks independently. Such risks are not only prevalent in widespread social media platforms~\cite{choi2023privacy,ma2025raising} but can also raise during task completion with smart wearable glasses~\cite{cai2025aiget} and on numerous self-logging devices~\cite{barker2024flexible}, potentially leading to diverse threats.

% Therefore, we advocate for improved privacy management methods. This call is partly driven by the inadequacy of current consent mechanisms~\cite{asthana2024know}. These mechanisms fail to inform users about inferential risks, much less enable granular control over such privacy-sensitive data. We propose exploring automated, content-aware fine-grained control methods~\cite{venugopalan2024aragorn} and leveraging them to address inferential privacy risks~\cite{zhang2024ghost}. For instance, this could involve performing inference using on-device models, presenting the inferred information to users, and allowing them to edit it via an interface.

\subsection{The Duality of Inference}

We followed Staab et al.~\cite{staab2023beyond} for deciding the personal attributes of the inference, where different information has different chances of being inferred. In the realm of recommendations, inferred user profiles can contain diverse aspects beyond our current scope~\cite{zhang2023chatgpt,wang2023survey}, and may benefit users in the form of providing personalized service~\cite{asthana2024know}. However, in our interview, most of the users mainly articulated the negative aspects of inferences rather than the positive aspects, citing it as a threat rather than benefits. This reflects participants' privacy calculus~\cite{laufer1977privacy} that the benefits seems minimal in comparison to the negative consequences, especially on social medias. This could be because that participants held an opaque models about the data usage before, therefore cannot easily think out beneficial usage. We acknowledged the duality of this inference, and deemed the detailed trade-offs of this inference as the future work. 

\subsection{Accuracy of Inference and their Effects}

Our interviews revealed users hold dichotomous perceptions of erroneous inferences from VLMs. One user group believes the model memorizes these errors, compelling them to make corrections to prevent future service degradation and viewing the inaccuracies as systemic flaws. In contrast, a second group dismisses incorrect inferences as inconsequential to their privacy and thus ignores them. Regardless of their mental model, users consistently reported that such errors induce confusion and cognitive load, fostering a perception of the system as ``broken''.

Fundamentally, VLM inferences are probabilistic estimations estimations based on limited cues and are thus inherently prone to inaccuracy and bias~\cite{staab2023beyond,zhang2024ghost}. These errors can cause harmful misrepresentations, such as stereotyping a user wearing a skirt as female. We argue that potential harm is not limited to accurate inference. Therefore, VLMs should refrain from making definitive but potentially incorrect claims. Instead, when inferential confidence is low, systems should be designed to express ambiguity or transparently solicit user clarification.

\subsection{Complexity and Burden of User-Driven Privacy Mitigation}

We found that users employ proactive content obfuscation methods to mitigate privacy risks. However, these user-driven mitigation strategies often prove insufficient. Many users propose simply avoiding the upload of ``risky'' videos altogether, which significantly compromises usability. For users, blurring individual objects within videos is cumbersome, especially given the current lack of robust video privacy editing tools. Prior anonymization techniques~\cite{venugopalan2024aragorn,zhang2024adanonymizer} and approaches to inferential privacy~\cite{zhang2025through} are predominantly image-based, failing to account for the dynamic nature of video and its associated challenges. Therefore, we advocate for improved methods that effectively balance user effort with user control and agency. This could involve direct user control or the adoption of automated method, where intelligent agents assist users by learning and applying their privacy preferences~\cite{zhang2025evaluating,zhang2025patient}.

% \subsection{The Inflexible Neglect of Inference Risks and its reason}
% Previous studies have predominantly focused on the risks associated with direct exposure of privacy information, which is about personal details appearing explicitly in photos or videos, such as address and identification number. However, our research suggests that inference privacy risks may be more severe than the risks of direct exposure. This is because the risks associated with direct exposure are more apparent, and users are typically able to recognize and avoid them. In contrast, the mechanisms underlying inference privacy risks are not easily understood or acknowledged by users, which leads them to focus primarily on direct exposure risks while overlooking other seemingly less critical information. In terms of mitigation strategies, users can address the risks of direct exposure by avoiding the inclusion of specific textual information in photos or videos. However, when it comes to inference privacy risks, users are not aware of the underlying mechanisms, making it difficult for them to take effective proactive measures. Therefore, we advocate for more transparent communication strategies that clearly outline the mechanisms of inference privacy in privacy management, in order to raise users' awareness of these types of privacy risks.

\subsection{Design Implications}

Our findings reveal a significant discrepancy between the inferential capabilities of VLMs and user perception. We therefore propose the following design implications, spanning interface design, system architecture, and platform governance, to foster a trustworthy and user-centric VLM ecosystem.

\textbf{Implication 1: Designing for inferential awareness.}
To mitigate user distress from unexpected and unsettlingly accurate inferences, systems must enhance inferential awareness. Instead of merely declaring what is inferred, interfaces should visually ground inferences in real-time by highlighting the specific visual evidence that contributes to a conclusion, such as cues indicating a location. This approach makes the inferential process transparent and scrutable to the user.

\textbf{Implication 2: Empowering user agency with proactive mitigation tools.} 
Our study shows that users perceive manual mitigation strategies as both ineffective and burdensome. To better empower users, systems should integrate proactive, AI-driven tools. Such tools would automatically detect and flag potential privacy risks in content prior to publication, offering users summarized assessments and one-click mitigation solutions, thereby reducing user burden~\cite{zhang2025through}.

\textbf{Implication 3: Shifting the burden through systemic safeguards and accountable governance.}
Users strongly advocate for shifting the onus of privacy protection from individual effort to platform-level responsibility. This necessitates systemic safeguards, including architecturally enforced and verifiable data lifecycle controls to rebuild trust. Furthermore, platforms must replace blanket, one-time consent models with granular consent mechanisms that require separate user approval for distinct classes of data inference, thereby aligning with users' preferences for selective sharing~\cite{venugopalan2024aragorn}.

\subsection{Limitations}

We acknowledge that our papers have three limitations. First, our recruitment is grounded in China, and the experiment involved solely Chinese participants with their experience primarily on Chinese platforms. Our recruitment is also biased towards university students. This may cause bias as Chinese users may create and upload different topics of videos on the social media platform. We envisioned a cross-cultural comparison of the inference risk as our future work. Second, although we provided vignettes for users and let them to reflect on their uploaded videos as examples, our qualitative study may subject to recall bias and desirability bias, which may not reflect the real opinions of participants. Thirdly, our experiment may not reflect the real capabilities of VLMs, as our experiment primarily involves interviews on participants about their mental models, strategies and expectations. Although the capability of VLMs are evolving, we believe the fundamental threats and risks are insightful for the future work.

\begin{acks}
This work was supported by the Natural Science Foundation of China under Grant No. 62472243 and 62132010.
\end{acks}
\bibliographystyle{ACM-Reference-Format}
\bibliography{sample-base}


\begin{thebibliography}{61}

%%% the Web address.

\ifx \showCODEN    \undefined \def \showCODEN     #1{\unskip}     \fi
\ifx \showISBNx    \undefined \def \showISBNx     #1{\unskip}     \fi
\ifx \showISBNxiii \undefined \def \showISBNxiii  #1{\unskip}     \fi
\ifx \showISSN     \undefined \def \showISSN      #1{\unskip}     \fi
\ifx \showLCCN     \undefined \def \showLCCN      #1{\unskip}     \fi
\ifx \shownote     \undefined \def \shownote      #1{#1}          \fi
\ifx \showarticletitle \undefined \def \showarticletitle #1{#1}   \fi
\ifx \showURL      \undefined \def \showURL       {\relax}        \fi
% invisible to TeX
\providecommand\bibfield[2]{#2}
\providecommand\bibinfo[2]{#2}
\providecommand\natexlab[1]{#1}
\providecommand\showeprint[2][]{arXiv:#2}

\bibitem[Aiordachioae and Vatavu(2019)]%
        {aiordachioae2019life}
\bibfield{author}{\bibinfo{person}{Adrian Aiordachioae} {and} \bibinfo{person}{Radu-Daniel Vatavu}.} \bibinfo{year}{2019}\natexlab{}.
\newblock \showarticletitle{Life-tags: a smartglasses-based system for recording and abstracting life with tag clouds}.
\newblock \bibinfo{journal}{\emph{Proceedings of the ACM on human-computer interaction}} \bibinfo{volume}{3}, \bibinfo{number}{EICS} (\bibinfo{year}{2019}), \bibinfo{pages}{1--22}.
\newblock


\bibitem[Anderson and Niu(2025)]%
        {anderson2025making}
\bibfield{author}{\bibinfo{person}{Torin Anderson} {and} \bibinfo{person}{Shuo Niu}.} \bibinfo{year}{2025}\natexlab{}.
\newblock \showarticletitle{Making AI-Enhanced Videos: Analyzing Generative AI Use Cases in YouTube Content Creation}. In \bibinfo{booktitle}{\emph{Proceedings of the Extended Abstracts of the CHI Conference on Human Factors in Computing Systems}}. \bibinfo{pages}{1--7}.
\newblock


\bibitem[Asthana et~al\mbox{.}(2024)]%
        {asthana2024know}
\bibfield{author}{\bibinfo{person}{Sumit Asthana}, \bibinfo{person}{Jane Im}, \bibinfo{person}{Zhe Chen}, {and} \bibinfo{person}{Nikola Banovic}.} \bibinfo{year}{2024}\natexlab{}.
\newblock \showarticletitle{" I know even if you don't tell me": Understanding Users' Privacy Preferences Regarding AI-based Inferences of Sensitive Information for Personalization}. In \bibinfo{booktitle}{\emph{Proceedings of the 2024 CHI Conference on Human Factors in Computing Systems}}. \bibinfo{pages}{1--21}.
\newblock


\bibitem[Bailey et~al\mbox{.}(2012)]%
        {bailey2012menlo}
\bibfield{author}{\bibinfo{person}{Michael Bailey}, \bibinfo{person}{David Dittrich}, \bibinfo{person}{Erin Kenneally}, {and} \bibinfo{person}{Doug Maughan}.} \bibinfo{year}{2012}\natexlab{}.
\newblock \showarticletitle{The menlo report}.
\newblock \bibinfo{journal}{\emph{IEEE Security \& Privacy}} \bibinfo{volume}{10}, \bibinfo{number}{2} (\bibinfo{year}{2012}), \bibinfo{pages}{71--75}.
\newblock


\bibitem[Barbosa et~al\mbox{.}(2020)]%
        {barbosa2020privacy}
\bibfield{author}{\bibinfo{person}{Nat{\~a}~M Barbosa}, \bibinfo{person}{Zhuohao Zhang}, {and} \bibinfo{person}{Yang Wang}.} \bibinfo{year}{2020}\natexlab{}.
\newblock \showarticletitle{Do Privacy and Security Matter to Everyone? Quantifying and Clustering $\{$User-Centric$\}$ Considerations About Smart Home Device Adoption}. In \bibinfo{booktitle}{\emph{Sixteenth Symposium on Usable Privacy and Security (SOUPS 2020)}}. \bibinfo{pages}{417--435}.
\newblock


\bibitem[Barker-Canler et~al\mbox{.}(2024)]%
        {barker2024flexible}
\bibfield{author}{\bibinfo{person}{Matthew Barker-Canler}, \bibinfo{person}{Daniel Gooch}, \bibinfo{person}{Janet Van Der~Linden}, {and} \bibinfo{person}{Marian Petre}.} \bibinfo{year}{2024}\natexlab{}.
\newblock \showarticletitle{Flexible minimalist self-tracking to support individual reflection}.
\newblock \bibinfo{journal}{\emph{ACM Transactions on Computer-Human Interaction}} \bibinfo{volume}{31}, \bibinfo{number}{3} (\bibinfo{year}{2024}), \bibinfo{pages}{1--35}.
\newblock


\bibitem[Beauchamp et~al\mbox{.}(2008)]%
        {beauchamp2008belmont}
\bibfield{author}{\bibinfo{person}{Tom~L Beauchamp} {et~al\mbox{.}}} \bibinfo{year}{2008}\natexlab{}.
\newblock \showarticletitle{The belmont report}.
\newblock \bibinfo{journal}{\emph{The Oxford textbook of clinical research ethics}} (\bibinfo{year}{2008}), \bibinfo{pages}{149--155}.
\newblock


\bibitem[Bhardwaj et~al\mbox{.}(2024)]%
        {bhardwaj2024focus}
\bibfield{author}{\bibinfo{person}{Divyanshu Bhardwaj}, \bibinfo{person}{Alexander Ponticello}, \bibinfo{person}{Shreya Tomar}, \bibinfo{person}{Adrian Dabrowski}, {and} \bibinfo{person}{Katharina Krombholz}.} \bibinfo{year}{2024}\natexlab{}.
\newblock \showarticletitle{In Focus, Out of Privacy: The Wearer's Perspective on the Privacy Dilemma of Camera Glasses}. In \bibinfo{booktitle}{\emph{Proceedings of the CHI Conference on Human Factors in Computing Systems}}. \bibinfo{pages}{1--18}.
\newblock


\bibitem[Bubeck et~al\mbox{.}(2023)]%
        {bubeck2023sparks}
\bibfield{author}{\bibinfo{person}{S{\'e}bastien Bubeck}, \bibinfo{person}{Varun Chandrasekaran}, \bibinfo{person}{Ronen Eldan}, \bibinfo{person}{Johannes Gehrke}, \bibinfo{person}{Eric Horvitz}, \bibinfo{person}{Ece Kamar}, \bibinfo{person}{Peter Lee}, \bibinfo{person}{Yin~Tat Lee}, \bibinfo{person}{Yuanzhi Li}, \bibinfo{person}{Scott Lundberg}, {et~al\mbox{.}}} \bibinfo{year}{2023}\natexlab{}.
\newblock \showarticletitle{Sparks of artificial general intelligence: Early experiments with gpt-4}.
\newblock \bibinfo{journal}{\emph{arXiv preprint arXiv:2303.12712}} (\bibinfo{year}{2023}).
\newblock


\bibitem[Cai et~al\mbox{.}(2025)]%
        {cai2025aiget}
\bibfield{author}{\bibinfo{person}{Runze Cai}, \bibinfo{person}{Nuwan Janaka}, \bibinfo{person}{Hyeongcheol Kim}, \bibinfo{person}{Yang Chen}, \bibinfo{person}{Shengdong Zhao}, \bibinfo{person}{Yun Huang}, {and} \bibinfo{person}{David Hsu}.} \bibinfo{year}{2025}\natexlab{}.
\newblock \showarticletitle{AiGet: Transforming Everyday Moments into Hidden Knowledge Discovery with AI Assistance on Smart Glasses}. In \bibinfo{booktitle}{\emph{Proceedings of the 2025 CHI Conference on Human Factors in Computing Systems}}. \bibinfo{pages}{1--26}.
\newblock


\bibitem[Castrill{\'o}n-Santana et~al\mbox{.}(2023)]%
        {castrillon2023evaluation}
\bibfield{author}{\bibinfo{person}{Modesto Castrill{\'o}n-Santana}, \bibinfo{person}{Elena S{\'a}nchez-Nielsen}, \bibinfo{person}{David Freire-Obreg{\'o}n}, \bibinfo{person}{Oliverio~J Santana}, \bibinfo{person}{Daniel Hern{\'a}ndez-Sosa}, {and} \bibinfo{person}{Javier Lorenzo-Navarro}.} \bibinfo{year}{2023}\natexlab{}.
\newblock \showarticletitle{Evaluation of a Visual Question Answering Architecture for Pedestrian Attribute Recognition}. In \bibinfo{booktitle}{\emph{International Conference on Computer Analysis of Images and Patterns}}. Springer, \bibinfo{pages}{13--22}.
\newblock


\bibitem[Cheng et~al\mbox{.}(2022)]%
        {cheng2022simple}
\bibfield{author}{\bibinfo{person}{Xinhua Cheng}, \bibinfo{person}{Mengxi Jia}, \bibinfo{person}{Qian Wang}, {and} \bibinfo{person}{Jian Zhang}.} \bibinfo{year}{2022}\natexlab{}.
\newblock \showarticletitle{A simple visual-textual baseline for pedestrian attribute recognition}.
\newblock \bibinfo{journal}{\emph{IEEE Transactions on Circuits and Systems for Video Technology}} \bibinfo{volume}{32}, \bibinfo{number}{10} (\bibinfo{year}{2022}), \bibinfo{pages}{6994--7004}.
\newblock


\bibitem[Choi(2023)]%
        {choi2023privacy}
\bibfield{author}{\bibinfo{person}{SoeYoon Choi}.} \bibinfo{year}{2023}\natexlab{}.
\newblock \showarticletitle{Privacy literacy on social media: Its predictors and outcomes}.
\newblock \bibinfo{journal}{\emph{International Journal of Human--Computer Interaction}} \bibinfo{volume}{39}, \bibinfo{number}{1} (\bibinfo{year}{2023}), \bibinfo{pages}{217--232}.
\newblock


\bibitem[Clarke and Braun(2014)]%
        {clarke2014thematic}
\bibfield{author}{\bibinfo{person}{Victoria Clarke} {and} \bibinfo{person}{Virginia Braun}.} \bibinfo{year}{2014}\natexlab{}.
\newblock \showarticletitle{Thematic analysis}.
\newblock In \bibinfo{booktitle}{\emph{Encyclopedia of critical psychology}}. \bibinfo{publisher}{Springer}, \bibinfo{pages}{1947--1952}.
\newblock


\bibitem[Denning et~al\mbox{.}(2014)]%
        {denning2014situ}
\bibfield{author}{\bibinfo{person}{Tamara Denning}, \bibinfo{person}{Zakariya Dehlawi}, {and} \bibinfo{person}{Tadayoshi Kohno}.} \bibinfo{year}{2014}\natexlab{}.
\newblock \showarticletitle{In situ with bystanders of augmented reality glasses: Perspectives on recording and privacy-mediating technologies}. In \bibinfo{booktitle}{\emph{Proceedings of the SIGCHI Conference on Human Factors in Computing Systems}}. \bibinfo{pages}{2377--2386}.
\newblock


\bibitem[Duddu and Boutet(2022)]%
        {duddu2022inferring}
\bibfield{author}{\bibinfo{person}{Vasisht Duddu} {and} \bibinfo{person}{Antoine Boutet}.} \bibinfo{year}{2022}\natexlab{}.
\newblock \showarticletitle{Inferring sensitive attributes from model explanations}. In \bibinfo{booktitle}{\emph{Proceedings of the 31st ACM International Conference on Information \& Knowledge Management}}. \bibinfo{pages}{416--425}.
\newblock


\bibitem[Eslami et~al\mbox{.}(2018)]%
        {eslami2018communicating}
\bibfield{author}{\bibinfo{person}{Motahhare Eslami}, \bibinfo{person}{Sneha~R Krishna~Kumaran}, \bibinfo{person}{Christian Sandvig}, {and} \bibinfo{person}{Karrie Karahalios}.} \bibinfo{year}{2018}\natexlab{}.
\newblock \showarticletitle{Communicating algorithmic process in online behavioral advertising}. In \bibinfo{booktitle}{\emph{Proceedings of the 2018 CHI conference on human factors in computing systems}}. \bibinfo{pages}{1--13}.
\newblock


\bibitem[F{\o}lstad et~al\mbox{.}(2018)]%
        {folstad2018makes}
\bibfield{author}{\bibinfo{person}{Asbj{\o}rn F{\o}lstad}, \bibinfo{person}{Cecilie~Bertinussen Nordheim}, {and} \bibinfo{person}{Cato~Alexander Bj{\o}rkli}.} \bibinfo{year}{2018}\natexlab{}.
\newblock \showarticletitle{What makes users trust a chatbot for customer service? An exploratory interview study}. In \bibinfo{booktitle}{\emph{Internet Science: 5th International Conference, INSCI 2018, St. Petersburg, Russia, October 24--26, 2018, Proceedings 5}}. Springer, \bibinfo{pages}{194--208}.
\newblock


\bibitem[Griffiths et~al\mbox{.}(2018)]%
        {griffiths2018privacy}
\bibfield{author}{\bibinfo{person}{Erin Griffiths}, \bibinfo{person}{Salah Assana}, {and} \bibinfo{person}{Kamin Whitehouse}.} \bibinfo{year}{2018}\natexlab{}.
\newblock \showarticletitle{Privacy-preserving image processing with binocular thermal cameras}.
\newblock \bibinfo{journal}{\emph{Proceedings of the ACM on Interactive, Mobile, Wearable and Ubiquitous Technologies}} \bibinfo{volume}{1}, \bibinfo{number}{4} (\bibinfo{year}{2018}), \bibinfo{pages}{1--25}.
\newblock


\bibitem[Harborth and Frik(2021)]%
        {harborth2021evaluating}
\bibfield{author}{\bibinfo{person}{David Harborth} {and} \bibinfo{person}{Alisa Frik}.} \bibinfo{year}{2021}\natexlab{}.
\newblock \showarticletitle{Evaluating and redefining smartphone permissions with contextualized justifications for mobile augmented reality apps}. In \bibinfo{booktitle}{\emph{Seventeenth Symposium on Usable Privacy and Security (SOUPS 2021)}}. \bibinfo{pages}{513--534}.
\newblock


\bibitem[Henne and Smith(2013)]%
        {henne2013awareness}
\bibfield{author}{\bibinfo{person}{Benjamin Henne} {and} \bibinfo{person}{Matthew Smith}.} \bibinfo{year}{2013}\natexlab{}.
\newblock \showarticletitle{Awareness about photos on the web and how privacy-privacy-tradeoffs could help}. In \bibinfo{booktitle}{\emph{Financial Cryptography and Data Security: FC 2013 Workshops, USEC and WAHC 2013, Okinawa, Japan, April 1, 2013, Revised Selected Papers 17}}. Springer, \bibinfo{pages}{131--148}.
\newblock


\bibitem[Hua et~al\mbox{.}(2024)]%
        {hua2024generative}
\bibfield{author}{\bibinfo{person}{Yiqing Hua}, \bibinfo{person}{Shuo Niu}, \bibinfo{person}{Jie Cai}, \bibinfo{person}{Lydia~B Chilton}, \bibinfo{person}{Hendrik Heuer}, {and} \bibinfo{person}{Donghee~Yvette Wohn}.} \bibinfo{year}{2024}\natexlab{}.
\newblock \showarticletitle{Generative AI in user-generated content}. In \bibinfo{booktitle}{\emph{Extended Abstracts of the CHI Conference on Human Factors in Computing Systems}}. \bibinfo{pages}{1--7}.
\newblock


\bibitem[Jung and Philipose(2014)]%
        {jung2014courteous}
\bibfield{author}{\bibinfo{person}{Jaeyeon Jung} {and} \bibinfo{person}{Matthai Philipose}.} \bibinfo{year}{2014}\natexlab{}.
\newblock \showarticletitle{Courteous glass}. In \bibinfo{booktitle}{\emph{Proceedings of the 2014 ACM international joint conference on pervasive and ubiquitous computing: Adjunct publication}}. \bibinfo{pages}{1307--1312}.
\newblock


\bibitem[Koelle et~al\mbox{.}(2018)]%
        {koelle2018your}
\bibfield{author}{\bibinfo{person}{Marion Koelle}, \bibinfo{person}{Swamy Ananthanarayan}, \bibinfo{person}{Simon Czupalla}, \bibinfo{person}{Wilko Heuten}, {and} \bibinfo{person}{Susanne Boll}.} \bibinfo{year}{2018}\natexlab{}.
\newblock \showarticletitle{Your smart glasses' camera bothers me! exploring opt-in and opt-out gestures for privacy mediation}. In \bibinfo{booktitle}{\emph{Proceedings of the 10th Nordic Conference on Human-Computer Interaction}}. \bibinfo{pages}{473--481}.
\newblock


\bibitem[Koelle et~al\mbox{.}(2015)]%
        {koelle2015don}
\bibfield{author}{\bibinfo{person}{Marion Koelle}, \bibinfo{person}{Matthias Kranz}, {and} \bibinfo{person}{Andreas M{\"o}ller}.} \bibinfo{year}{2015}\natexlab{}.
\newblock \showarticletitle{Don't look at me that way! Understanding user attitudes towards data glasses usage}. In \bibinfo{booktitle}{\emph{Proceedings of the 17th international conference on human-computer interaction with mobile devices and services}}. \bibinfo{pages}{362--372}.
\newblock


\bibitem[Kozyreva et~al\mbox{.}(2021)]%
        {kozyreva2021public}
\bibfield{author}{\bibinfo{person}{Anastasia Kozyreva}, \bibinfo{person}{Philipp Lorenz-Spreen}, \bibinfo{person}{Ralph Hertwig}, \bibinfo{person}{Stephan Lewandowsky}, {and} \bibinfo{person}{Stefan~M Herzog}.} \bibinfo{year}{2021}\natexlab{}.
\newblock \showarticletitle{Public attitudes towards algorithmic personalization and use of personal data online: Evidence from Germany, Great Britain, and the United States}.
\newblock \bibinfo{journal}{\emph{Humanities and Social Sciences Communications}} \bibinfo{volume}{8}, \bibinfo{number}{1} (\bibinfo{year}{2021}), \bibinfo{pages}{1--11}.
\newblock


\bibitem[Lau et~al\mbox{.}(2018)]%
        {lau2018alexa}
\bibfield{author}{\bibinfo{person}{Josephine Lau}, \bibinfo{person}{Benjamin Zimmerman}, {and} \bibinfo{person}{Florian Schaub}.} \bibinfo{year}{2018}\natexlab{}.
\newblock \showarticletitle{Alexa, are you listening? Privacy perceptions, concerns and privacy-seeking behaviors with smart speakers}.
\newblock \bibinfo{journal}{\emph{Proceedings of the ACM on human-computer interaction}} \bibinfo{volume}{2}, \bibinfo{number}{CSCW} (\bibinfo{year}{2018}), \bibinfo{pages}{1--31}.
\newblock


\bibitem[Laufer and Wolfe(1977)]%
        {laufer1977privacy}
\bibfield{author}{\bibinfo{person}{Robert~S Laufer} {and} \bibinfo{person}{Maxine Wolfe}.} \bibinfo{year}{1977}\natexlab{}.
\newblock \showarticletitle{Privacy as a concept and a social issue: A multidimensional developmental theory}.
\newblock \bibinfo{journal}{\emph{Journal of social Issues}} \bibinfo{volume}{33}, \bibinfo{number}{3} (\bibinfo{year}{1977}), \bibinfo{pages}{22--42}.
\newblock


\bibitem[Liu et~al\mbox{.}(2022)]%
        {liu2022effects}
\bibfield{author}{\bibinfo{person}{Yu-li Liu}, \bibinfo{person}{Wenjia Yan}, \bibinfo{person}{Bo Hu}, \bibinfo{person}{Zhuoyang Li}, {and} \bibinfo{person}{Yik~Ling Lai}.} \bibinfo{year}{2022}\natexlab{}.
\newblock \showarticletitle{Effects of personalization and source expertise on users’ health beliefs and usage intention toward health chatbots: Evidence from an online experiment}.
\newblock \bibinfo{journal}{\emph{Digital Health}}  \bibinfo{volume}{8} (\bibinfo{year}{2022}), \bibinfo{pages}{20552076221129718}.
\newblock


\bibitem[Ma et~al\mbox{.}(2025)]%
        {ma2025raising}
\bibfield{author}{\bibinfo{person}{Ying Ma}, \bibinfo{person}{Shiquan Zhang}, \bibinfo{person}{Dongju Yang}, \bibinfo{person}{Zhanna Sarsenbayeva}, \bibinfo{person}{Jarrod Knibbe}, {and} \bibinfo{person}{Jorge Goncalves}.} \bibinfo{year}{2025}\natexlab{}.
\newblock \showarticletitle{Raising Awareness of Location Information Vulnerabilities in Social Media Photos using LLMs}. In \bibinfo{booktitle}{\emph{Proceedings of the 2025 CHI Conference on Human Factors in Computing Systems}}. \bibinfo{pages}{1--14}.
\newblock


\bibitem[Malkin et~al\mbox{.}(2022)]%
        {malkin2022runtime}
\bibfield{author}{\bibinfo{person}{Nathan Malkin}, \bibinfo{person}{David Wagner}, {and} \bibinfo{person}{Serge Egelman}.} \bibinfo{year}{2022}\natexlab{}.
\newblock \showarticletitle{Runtime permissions for privacy in proactive intelligent assistants}. In \bibinfo{booktitle}{\emph{Eighteenth Symposium on Usable Privacy and Security (SOUPS 2022)}}. \bibinfo{pages}{633--651}.
\newblock


\bibitem[Mehnaz et~al\mbox{.}(2022)]%
        {mehnaz2022your}
\bibfield{author}{\bibinfo{person}{Shagufta Mehnaz}, \bibinfo{person}{Sayanton~V Dibbo}, \bibinfo{person}{Ehsanul Kabir}, \bibinfo{person}{Ninghui Li}, {and} \bibinfo{person}{Elisa Bertino}.} \bibinfo{year}{2022}\natexlab{}.
\newblock \showarticletitle{Are your sensitive attributes private? novel model inversion attribute inference attacks on classification models}. In \bibinfo{booktitle}{\emph{31st USENIX Security Symposium (USENIX Security 22)}}. \bibinfo{pages}{4579--4596}.
\newblock


\bibitem[Milne et~al\mbox{.}(2017)]%
        {milne2017information}
\bibfield{author}{\bibinfo{person}{George~R Milne}, \bibinfo{person}{George Pettinico}, \bibinfo{person}{Fatima~M Hajjat}, {and} \bibinfo{person}{Ereni Markos}.} \bibinfo{year}{2017}\natexlab{}.
\newblock \showarticletitle{Information sensitivity typology: Mapping the degree and type of risk consumers perceive in personal data sharing}.
\newblock \bibinfo{journal}{\emph{Journal of Consumer Affairs}} \bibinfo{volume}{51}, \bibinfo{number}{1} (\bibinfo{year}{2017}), \bibinfo{pages}{133--161}.
\newblock


\bibitem[Nimmo et~al\mbox{.}(2024)]%
        {nimmo2024user}
\bibfield{author}{\bibinfo{person}{Robert Nimmo}, \bibinfo{person}{Marios Constantinides}, \bibinfo{person}{Ke Zhou}, \bibinfo{person}{Daniele Quercia}, {and} \bibinfo{person}{Simone Stumpf}.} \bibinfo{year}{2024}\natexlab{}.
\newblock \showarticletitle{User Characteristics in Explainable AI: The Rabbit Hole of Personalization?}. In \bibinfo{booktitle}{\emph{Proceedings of the 2024 CHI Conference on Human Factors in Computing Systems}}. \bibinfo{pages}{1--13}.
\newblock


\bibitem[Olson et~al\mbox{.}(2005)]%
        {olson2005study}
\bibfield{author}{\bibinfo{person}{Judith~S Olson}, \bibinfo{person}{Jonathan Grudin}, {and} \bibinfo{person}{Eric Horvitz}.} \bibinfo{year}{2005}\natexlab{}.
\newblock \showarticletitle{A study of preferences for sharing and privacy}. In \bibinfo{booktitle}{\emph{CHI'05 extended abstracts on Human factors in computing systems}}. \bibinfo{pages}{1985--1988}.
\newblock


\bibitem[Orekondy et~al\mbox{.}(2018)]%
        {orekondy2018connecting}
\bibfield{author}{\bibinfo{person}{Tribhuvanesh Orekondy}, \bibinfo{person}{Mario Fritz}, {and} \bibinfo{person}{Bernt Schiele}.} \bibinfo{year}{2018}\natexlab{}.
\newblock \showarticletitle{Connecting pixels to privacy and utility: Automatic redaction of private information in images}. In \bibinfo{booktitle}{\emph{Proceedings of the IEEE Conference on Computer Vision and Pattern Recognition}}. \bibinfo{pages}{8466--8475}.
\newblock


\bibitem[Orekondy et~al\mbox{.}(2017)]%
        {orekondy2017towards}
\bibfield{author}{\bibinfo{person}{Tribhuvanesh Orekondy}, \bibinfo{person}{Bernt Schiele}, {and} \bibinfo{person}{Mario Fritz}.} \bibinfo{year}{2017}\natexlab{}.
\newblock \showarticletitle{Towards a visual privacy advisor: Understanding and predicting privacy risks in images}. In \bibinfo{booktitle}{\emph{Proceedings of the IEEE international conference on computer vision}}. \bibinfo{pages}{3686--3695}.
\newblock


\bibitem[Rader et~al\mbox{.}(2020)]%
        {rader2020have}
\bibfield{author}{\bibinfo{person}{Emilee Rader}, \bibinfo{person}{Samantha Hautea}, {and} \bibinfo{person}{Anjali Munasinghe}.} \bibinfo{year}{2020}\natexlab{}.
\newblock \showarticletitle{" I Have a Narrow Thought Process": Constraints on Explanations Connecting Inferences and $\{$Self-Perceptions$\}$}. In \bibinfo{booktitle}{\emph{Sixteenth Symposium on Usable Privacy and Security (SOUPS 2020)}}. \bibinfo{pages}{457--488}.
\newblock


\bibitem[Reardon et~al\mbox{.}(2019)]%
        {reardon201950}
\bibfield{author}{\bibinfo{person}{Joel Reardon}, \bibinfo{person}{{\'A}lvaro Feal}, \bibinfo{person}{Primal Wijesekera}, \bibinfo{person}{Amit Elazari~Bar On}, \bibinfo{person}{Narseo Vallina-Rodriguez}, {and} \bibinfo{person}{Serge Egelman}.} \bibinfo{year}{2019}\natexlab{}.
\newblock \showarticletitle{50 ways to leak your data: An exploration of apps' circumvention of the android permissions system}. In \bibinfo{booktitle}{\emph{28th USENIX security symposium (USENIX security 19)}}. \bibinfo{pages}{603--620}.
\newblock


\bibitem[Shi et~al\mbox{.}(2021)]%
        {shi2021face}
\bibfield{author}{\bibinfo{person}{Cong Shi}, \bibinfo{person}{Xiangyu Xu}, \bibinfo{person}{Tianfang Zhang}, \bibinfo{person}{Payton Walker}, \bibinfo{person}{Yi Wu}, \bibinfo{person}{Jian Liu}, \bibinfo{person}{Nitesh Saxena}, \bibinfo{person}{Yingying Chen}, {and} \bibinfo{person}{Jiadi Yu}.} \bibinfo{year}{2021}\natexlab{}.
\newblock \showarticletitle{Face-Mic: inferring live speech and speaker identity via subtle facial dynamics captured by AR/VR motion sensors}. In \bibinfo{booktitle}{\emph{Proceedings of the 27th Annual International Conference on Mobile Computing and Networking}}. \bibinfo{pages}{478--490}.
\newblock


\bibitem[Staab et~al\mbox{.}({[n.\,d.]})]%
        {staab2023beyond}
\bibfield{author}{\bibinfo{person}{Robin Staab}, \bibinfo{person}{Mark Vero}, \bibinfo{person}{Mislav Balunovic}, {and} \bibinfo{person}{Martin Vechev}.} \bibinfo{year}{[n.\,d.]}\natexlab{}.
\newblock \showarticletitle{Beyond Memorization: Violating Privacy via Inference with Large Language Models}. In \bibinfo{booktitle}{\emph{The Twelfth International Conference on Learning Representations}}.
\newblock


\bibitem[Steil et~al\mbox{.}(2019)]%
        {steil2019privaceye}
\bibfield{author}{\bibinfo{person}{Julian Steil}, \bibinfo{person}{Marion Koelle}, \bibinfo{person}{Wilko Heuten}, \bibinfo{person}{Susanne Boll}, {and} \bibinfo{person}{Andreas Bulling}.} \bibinfo{year}{2019}\natexlab{}.
\newblock \showarticletitle{Privaceye: privacy-preserving head-mounted eye tracking using egocentric scene image and eye movement features}. In \bibinfo{booktitle}{\emph{Proceedings of the 11th ACM symposium on eye tracking research \& applications}}. \bibinfo{pages}{1--10}.
\newblock


\bibitem[T{\"o}mek{\c{c}}e et~al\mbox{.}(2024)]%
        {tomekcce2024private}
\bibfield{author}{\bibinfo{person}{Batuhan T{\"o}mek{\c{c}}e}, \bibinfo{person}{Mark Vero}, \bibinfo{person}{Robin Staab}, {and} \bibinfo{person}{Martin Vechev}.} \bibinfo{year}{2024}\natexlab{}.
\newblock \showarticletitle{Private Attribute Inference from Images with Vision-Language Models}.
\newblock \bibinfo{journal}{\emph{arXiv preprint arXiv:2404.10618}} (\bibinfo{year}{2024}).
\newblock


\bibitem[Venugopalan et~al\mbox{.}(2024)]%
        {venugopalan2024aragorn}
\bibfield{author}{\bibinfo{person}{Hari Venugopalan}, \bibinfo{person}{Zainul~Abi Din}, \bibinfo{person}{Trevor Carpenter}, \bibinfo{person}{Jason Lowe-Power}, \bibinfo{person}{Samuel~T King}, {and} \bibinfo{person}{Zubair Shafiq}.} \bibinfo{year}{2024}\natexlab{}.
\newblock \showarticletitle{Aragorn: A privacy-enhancing system for mobile cameras}.
\newblock \bibinfo{journal}{\emph{Proceedings of the ACM on Interactive, Mobile, Wearable and Ubiquitous Technologies}} \bibinfo{volume}{7}, \bibinfo{number}{4} (\bibinfo{year}{2024}), \bibinfo{pages}{1--31}.
\newblock


\bibitem[Vimalkumar et~al\mbox{.}(2021)]%
        {vimalkumar2021okay}
\bibfield{author}{\bibinfo{person}{M Vimalkumar}, \bibinfo{person}{Sujeet~Kumar Sharma}, \bibinfo{person}{Jang~Bahadur Singh}, {and} \bibinfo{person}{Yogesh~K Dwivedi}.} \bibinfo{year}{2021}\natexlab{}.
\newblock \showarticletitle{‘Okay google, what about my privacy?’: User's privacy perceptions and acceptance of voice based digital assistants}.
\newblock \bibinfo{journal}{\emph{Computers in Human Behavior}}  \bibinfo{volume}{120} (\bibinfo{year}{2021}), \bibinfo{pages}{106763}.
\newblock


\bibitem[Wang et~al\mbox{.}(2024)]%
        {wang2024pedestrian}
\bibfield{author}{\bibinfo{person}{Xiao Wang}, \bibinfo{person}{Jiandong Jin}, \bibinfo{person}{Chenglong Li}, \bibinfo{person}{Jin Tang}, \bibinfo{person}{Cheng Zhang}, {and} \bibinfo{person}{Wei Wang}.} \bibinfo{year}{2024}\natexlab{}.
\newblock \showarticletitle{Pedestrian attribute recognition via clip based prompt vision-language fusion}.
\newblock \bibinfo{journal}{\emph{IEEE Transactions on Circuits and Systems for Video Technology}} (\bibinfo{year}{2024}).
\newblock


\bibitem[Wang et~al\mbox{.}({[n.\,d.]})]%
        {wangself}
\bibfield{author}{\bibinfo{person}{Xuezhi Wang}, \bibinfo{person}{Jason Wei}, \bibinfo{person}{Dale Schuurmans}, \bibinfo{person}{Quoc~V Le}, \bibinfo{person}{Ed~H Chi}, \bibinfo{person}{Sharan Narang}, \bibinfo{person}{Aakanksha Chowdhery}, {and} \bibinfo{person}{Denny Zhou}.} \bibinfo{year}{[n.\,d.]}\natexlab{}.
\newblock \showarticletitle{Self-Consistency Improves Chain of Thought Reasoning in Language Models}. In \bibinfo{booktitle}{\emph{The Eleventh International Conference on Learning Representations}}.
\newblock


\bibitem[Wang et~al\mbox{.}(2023)]%
        {wang2023survey}
\bibfield{author}{\bibinfo{person}{Yifan Wang}, \bibinfo{person}{Weizhi Ma}, \bibinfo{person}{Min Zhang}, \bibinfo{person}{Yiqun Liu}, {and} \bibinfo{person}{Shaoping Ma}.} \bibinfo{year}{2023}\natexlab{}.
\newblock \showarticletitle{A survey on the fairness of recommender systems}.
\newblock \bibinfo{journal}{\emph{ACM Transactions on Information Systems}} \bibinfo{volume}{41}, \bibinfo{number}{3} (\bibinfo{year}{2023}), \bibinfo{pages}{1--43}.
\newblock


\bibitem[Warshaw et~al\mbox{.}(2016)]%
        {warshaw2016intuitions}
\bibfield{author}{\bibinfo{person}{Jeffrey Warshaw}, \bibinfo{person}{Nina Taft}, {and} \bibinfo{person}{Allison Woodruff}.} \bibinfo{year}{2016}\natexlab{}.
\newblock \showarticletitle{Intuitions, analytics, and killing ants: inference literacy of high school-educated adults in the $\{$US$\}$}. In \bibinfo{booktitle}{\emph{Twelfth Symposium on Usable Privacy and Security (SOUPS 2016)}}. \bibinfo{pages}{271--285}.
\newblock


\bibitem[Weinshel et~al\mbox{.}(2019)]%
        {weinshel2019oh}
\bibfield{author}{\bibinfo{person}{Ben Weinshel}, \bibinfo{person}{Miranda Wei}, \bibinfo{person}{Mainack Mondal}, \bibinfo{person}{Euirim Choi}, \bibinfo{person}{Shawn Shan}, \bibinfo{person}{Claire Dolin}, \bibinfo{person}{Michelle~L Mazurek}, {and} \bibinfo{person}{Blase Ur}.} \bibinfo{year}{2019}\natexlab{}.
\newblock \showarticletitle{Oh, the places you've been! User reactions to longitudinal transparency about third-party web tracking and inferencing}. In \bibinfo{booktitle}{\emph{Proceedings of the 2019 ACM SIGSAC Conference on Computer and Communications Security}}. \bibinfo{pages}{149--166}.
\newblock


\bibitem[Wills and Zeljkovic(2011)]%
        {wills2011personalized}
\bibfield{author}{\bibinfo{person}{Craig~E Wills} {and} \bibinfo{person}{Mihajlo Zeljkovic}.} \bibinfo{year}{2011}\natexlab{}.
\newblock \showarticletitle{A personalized approach to web privacy: awareness, attitudes and actions}.
\newblock \bibinfo{journal}{\emph{Information Management \& Computer Security}} \bibinfo{volume}{19}, \bibinfo{number}{1} (\bibinfo{year}{2011}), \bibinfo{pages}{53--73}.
\newblock


\bibitem[Wu et~al\mbox{.}(2020)]%
        {wu2020privacy}
\bibfield{author}{\bibinfo{person}{Zhenyu Wu}, \bibinfo{person}{Haotao Wang}, \bibinfo{person}{Zhaowen Wang}, \bibinfo{person}{Hailin Jin}, {and} \bibinfo{person}{Zhangyang Wang}.} \bibinfo{year}{2020}\natexlab{}.
\newblock \showarticletitle{Privacy-preserving deep action recognition: An adversarial learning framework and a new dataset}.
\newblock \bibinfo{journal}{\emph{IEEE Transactions on Pattern Analysis and Machine Intelligence}} \bibinfo{volume}{44}, \bibinfo{number}{4} (\bibinfo{year}{2020}), \bibinfo{pages}{2126--2139}.
\newblock


\bibitem[Yao et~al\mbox{.}(2019)]%
        {yao2019privacy}
\bibfield{author}{\bibinfo{person}{Yaxing Yao}, \bibinfo{person}{Justin~Reed Basdeo}, \bibinfo{person}{Oriana~Rosata Mcdonough}, {and} \bibinfo{person}{Yang Wang}.} \bibinfo{year}{2019}\natexlab{}.
\newblock \showarticletitle{Privacy perceptions and designs of bystanders in smart homes}.
\newblock \bibinfo{journal}{\emph{Proceedings of the ACM on Human-Computer Interaction}} \bibinfo{volume}{3}, \bibinfo{number}{CSCW} (\bibinfo{year}{2019}), \bibinfo{pages}{1--24}.
\newblock


\bibitem[Yao et~al\mbox{.}(2017)]%
        {yao2017folk}
\bibfield{author}{\bibinfo{person}{Yaxing Yao}, \bibinfo{person}{Davide Lo~Re}, {and} \bibinfo{person}{Yang Wang}.} \bibinfo{year}{2017}\natexlab{}.
\newblock \showarticletitle{Folk models of online behavioral advertising}. In \bibinfo{booktitle}{\emph{Proceedings of the 2017 ACM Conference on Computer Supported Cooperative Work and Social Computing}}. \bibinfo{pages}{1957--1969}.
\newblock


\bibitem[Zhang et~al\mbox{.}(2023)]%
        {zhang2023chatgpt}
\bibfield{author}{\bibinfo{person}{Jizhi Zhang}, \bibinfo{person}{Keqin Bao}, \bibinfo{person}{Yang Zhang}, \bibinfo{person}{Wenjie Wang}, \bibinfo{person}{Fuli Feng}, {and} \bibinfo{person}{Xiangnan He}.} \bibinfo{year}{2023}\natexlab{}.
\newblock \showarticletitle{Is chatgpt fair for recommendation? evaluating fairness in large language model recommendation}. In \bibinfo{booktitle}{\emph{Proceedings of the 17th ACM Conference on Recommender Systems}}. \bibinfo{pages}{993--999}.
\newblock


\bibitem[Zhang and Li(2024)]%
        {zhang2024confrontation}
\bibfield{author}{\bibinfo{person}{Shuning Zhang} {and} \bibinfo{person}{Shixuan Li}.} \bibinfo{year}{2024}\natexlab{}.
\newblock \showarticletitle{" Confrontation or Acceptance": Understanding Novice Visual Artists' Perception towards AI-assisted Art Creation}.
\newblock \bibinfo{journal}{\emph{arXiv preprint arXiv:2410.14925}} (\bibinfo{year}{2024}).
\newblock


\bibitem[Zhang et~al\mbox{.}(2025b)]%
        {zhang2025patient}
\bibfield{author}{\bibinfo{person}{Shuning Zhang}, \bibinfo{person}{Ying Ma}, \bibinfo{person}{YongquanOwen' Hu}, \bibinfo{person}{Ting Dang}, \bibinfo{person}{Hong Jia}, \bibinfo{person}{Xin Yi}, {and} \bibinfo{person}{Hewu Li}.} \bibinfo{year}{2025}\natexlab{b}.
\newblock \showarticletitle{From Patient Burdens to User Agency: Designing for Real-Time Protection Support in Online Health Consultations}.
\newblock \bibinfo{journal}{\emph{arXiv preprint arXiv:2508.00328}} (\bibinfo{year}{2025}).
\newblock


\bibitem[Zhang et~al\mbox{.}(2025c)]%
        {zhang2025evaluating}
\bibfield{author}{\bibinfo{person}{Shuning Zhang}, \bibinfo{person}{Ying Ma}, \bibinfo{person}{Xin Yi}, {and} \bibinfo{person}{Hewu Li}.} \bibinfo{year}{2025}\natexlab{c}.
\newblock \showarticletitle{Evaluating the Efficacy of Large Language Models for Generating Fine-Grained Visual Privacy Policies in Homes}.
\newblock \bibinfo{journal}{\emph{arXiv preprint arXiv:2508.00321}} (\bibinfo{year}{2025}).
\newblock


\bibitem[Zhang et~al\mbox{.}(2024a)]%
        {zhang2024ghost}
\bibfield{author}{\bibinfo{person}{Shuning Zhang}, \bibinfo{person}{Lyumanshan Ye}, \bibinfo{person}{Xin Yi}, \bibinfo{person}{Jingyu Tang}, \bibinfo{person}{Bo Shui}, \bibinfo{person}{Haobin Xing}, \bibinfo{person}{Pengfei Liu}, {and} \bibinfo{person}{Hewu Li}.} \bibinfo{year}{2024}\natexlab{a}.
\newblock \showarticletitle{" Ghost of the past": identifying and resolving privacy leakage from LLM's memory through proactive user interaction}.
\newblock \bibinfo{journal}{\emph{arXiv preprint arXiv:2410.14931}} (\bibinfo{year}{2024}).
\newblock


\bibitem[Zhang et~al\mbox{.}(2024b)]%
        {zhang2024adanonymizer}
\bibfield{author}{\bibinfo{person}{Shuning Zhang}, \bibinfo{person}{Xin Yi}, \bibinfo{person}{Haobin Xing}, \bibinfo{person}{Lyumanshan Ye}, \bibinfo{person}{Yongquan Hu}, {and} \bibinfo{person}{Hewu Li}.} \bibinfo{year}{2024}\natexlab{b}.
\newblock \showarticletitle{Adanonymizer: Interactively Navigating and Balancing the Duality of Privacy and Output Performance in Human-LLM Interaction}.
\newblock \bibinfo{journal}{\emph{arXiv preprint arXiv:2410.15044}} (\bibinfo{year}{2024}).
\newblock


\bibitem[Zhang et~al\mbox{.}(2025a)]%
        {zhang2025through}
\bibfield{author}{\bibinfo{person}{Ziyang Zhang}, \bibinfo{person}{Chong Bao}, \bibinfo{person}{Xiaokun Pan}, \bibinfo{person}{Chia-Ming Chang}, \bibinfo{person}{Takeo Igarashi}, {and} \bibinfo{person}{Guofeng Zhang}.} \bibinfo{year}{2025}\natexlab{a}.
\newblock \showarticletitle{Through the Lens of Privacy: Exploring Privacy Protection in Vision-Language Model Interactions on Smart Glasses}. In \bibinfo{booktitle}{\emph{Proceedings of the Extended Abstracts of the CHI Conference on Human Factors in Computing Systems}}. \bibinfo{pages}{1--8}.
\newblock


\end{thebibliography}

\appendix

\section{Ethics Considerations}

We acknowledged that our paper may have ethical concerns. We meticulously designed the study and considered all aspects related to the ethics. We followed Menlo Report~\cite{bailey2012menlo} and Belmont Report~\cite{beauchamp2008belmont} in designing the experiment, and our study got the approval of our institution's Institutional Review Board (IRB). Our study aims to understand users' perceptions towards VLM's inference on users' videos uploaded to the social media, and the potential privacy implications. Our results provided implications for designing ethical systems and platforms, which could benefit end-users. Before the study, we provided users' the user consent, and got the acceptance before conducting the study. We informed participants that our study is completely voluntary, and participants could quit the study at any time without penalty, and did not need to provide any explanation. We compensated participants appropriately according to the local wage standard. 

\section{Interview Script}

Our interview protocol was designed to explore participants’ perception with AI-generated private attribute inference, their mitigation strategies, and the effectiveness of their methods. The interview includes several sequentially progressive research questions. The original
interview script was conducted in simplified Chinese to ensure
natural communication with participants, and has been translated
into English for presentation in this paper.

\subsection{Research Questions}
\begin{itemize}
    \item Have you ever used a large model with visual capabilities? If so, have you ever felt that these models sometimes infer unexpected information from images? If so, can you give an example that left a strong impression on you?
    \item Do you think a large model can infer the aforementioned information (e.g., age range, gender) from videos you have previously filmed? If so, how do you think it makes these inferences? If not, why?
    \item Do you think if someone else filmed such a video (like the one you previously provided), it could infer the corresponding privacy attributes of the video creator (e.g., age range, gender)? If so, how would you infer this? If not, what conditions do you think are missing?
    \item Do you think this poses any negative risks to you? Specifically, what would you be concerned about?
    \item In what tasks or situations do you think the situation described above could specifically invade your privacy? Can you give an example that comes to your mind, or one that you think is the most relevant?
    \item In this example, what measures would you currently take to avoid such risks? Or, did you not realize these risks before? Can you describe the methods you are now taking?
    \item Do you think the methods you have taken are effective? Why?
    \item What methods do you think could be implemented in the future, such as interface-based or reminder box-based solutions, to avoid current privacy risks?
\end{itemize}

\section{User Consent}

We are a research group from XXX University, investigating on users' perception of AI-generated private attribute inference. The inference is mainly based on visual data that users upload to the model, especially VLMs, which can be analyzed and used to infer personal information of users, leading to privacy risks. Our study's focus is to understand your perception on the private inference of AI, your concerns of privacy risks caused by this, as well as your mitigation strategies and their effectiveness. 

The interview would take approximately 30-60 minutes depending on its content, and would be audio recorded, and transcribed for academic analysis and publication. We would not use your material including the audio and transcribed text for any other usage than outlined above. Your participation is completely voluntary, and you has the right to withdraw at any time without penalty or explanation. Your data would be kept confidential and anonymized before processing. If you complete the experiment, you could get a compensation of 90RMB per hour. 

If you have any other questions, you could contact XXXX (Email: XXXX) for further clarification.

\end{document}